\documentclass[gmd, manuscript]{copernicus}

\begin{document}

\nolinenumbers

\newcommand{\commShareFour}{33}
\newcommand{\coreMsIceAdd}{15}
\newcommand{\coreMsMixAdd}{15}
\newcommand{\coreMsOce}{55}
\newcommand{\coreParEffFour}{67}
\newcommand{\coreSYPD}{58}
\newcommand{\coreSYPDCPU}{0.9}
\newcommand{\coreSYPDOneGPU}{21}
\newcommand{\coreSYPDProd}{58}
\newcommand{\coreSYPDTwoNode}{67}
\newcommand{\coreStep}{0.085}
\newcommand{\coreStepCPU}{5.76}
\newcommand{\coreStepOneGPU}{0.229}
\newcommand{\coreStepProd}{0.085}
\newcommand{\coreStepTwoNode}{0.074}
\newcommand{\darsEffSixteen}{--}
\newcommand{\darsEffSixtyFour}{63}
\newcommand{\darsEffThirtyTwo}{83}
\newcommand{\darsSYPD}{1.46}
\newcommand{\darsStep}{0.45}
\newcommand{\darsStepOneTwentyEight}{0.19}
\newcommand{\darsStepSixtyFour}{0.22}
\newcommand{\forcaEffThirtyTwo}{75}
\newcommand{\forcaSYPD}{1.69}
\newcommand{\forcaStep}{0.39}
\newcommand{\forcaStepOneTwentyEight}{0.23}
\newcommand{\forcaStepSixtyFour}{0.23}
\newcommand{\ghRatioCoreEight}{1.61}
\newcommand{\ghRatioCoreFour}{1.96}
\newcommand{\ghRatioCoreOne}{2.46}
\newcommand{\ghRatioCoreTwo}{2.28}
\newcommand{\ghRatioDarsOneTwentyEight}{1.37}
\newcommand{\ghRatioDarsSixteen}{2.46}
\newcommand{\ghRatioDarsSixtyFour}{1.80}
\newcommand{\ghRatioDarsThirtyTwo}{2.04}
\newcommand{\ghRatioFarcEight}{1.27}
\newcommand{\ghRatioFarcFour}{1.93}
\newcommand{\ghRatioFarcSixteen}{0.90}
\newcommand{\ghRatioNgOneTwentyEight}{1.47}
\newcommand{\ghRatioNgSixtyFour}{1.86}
\newcommand{\ghRatioNgThirtyTwo}{2.38}
\newcommand{\ghStepCoreEight}{0.046}
\newcommand{\ghStepCoreFour}{0.044}
\newcommand{\ghStepCoreOne}{0.093}
\newcommand{\ghStepCoreSixteen}{0.061}
\newcommand{\ghStepCoreTwo}{0.064}
\newcommand{\ghStepDarsOneTwentyEight}{0.138}
\newcommand{\ghStepDarsSixteen}{0.183}
\newcommand{\ghStepDarsSixtyFour}{0.121}
\newcommand{\ghStepDarsThirtyTwo}{0.133}
\newcommand{\ghStepFarcEight}{0.156}
\newcommand{\ghStepFarcFour}{0.146}
\newcommand{\ghStepFarcSixteen}{0.170}
\newcommand{\ghStepFarcSixtyFour}{0.248}
\newcommand{\ghStepFarcThirtyTwo}{0.170}
\newcommand{\ghStepNgOneTwentyEight}{0.236}
\newcommand{\ghStepNgSixtyFour}{0.236}
\newcommand{\ghStepNgThirtyTwo}{0.317}
\newcommand{\ghStepNgTwoFiftySix}{0.248}
\newcommand{\ghSypdCoreEight}{106.9}
\newcommand{\ghSypdCoreFour}{112.8}
\newcommand{\ghSypdCoreFourRound}{113}
\newcommand{\ghSypdCoreOne}{52.9}
\newcommand{\ghSypdCoreSixteen}{81.1}
\newcommand{\ghSypdCoreTwo}{76.9}
\newcommand{\ghSypdDarsOneTwentyEight}{4.78}
\newcommand{\ghSypdDarsSixteen}{3.59}
\newcommand{\ghSypdDarsSixtyFour}{5.43}
\newcommand{\ghSypdDarsThirtyTwo}{4.93}
\newcommand{\ghSypdFarcEight}{21.0}
\newcommand{\ghSypdFarcFour}{22.4}
\newcommand{\ghSypdFarcSixteen}{19.3}
\newcommand{\ghSypdFarcSixtyFour}{13.3}
\newcommand{\ghSypdFarcThirtyTwo}{19.3}
\newcommand{\ghSypdNgOneTwentyEight}{2.78}
\newcommand{\ghSypdNgSixtyFour}{2.79}
\newcommand{\ghSypdNgThirtyTwo}{2.07}
\newcommand{\ghSypdNgTwoFiftySix}{2.65}
\newcommand{\hindcastMonths}{744}
\newcommand{\hindcastYears}{1958--2019}
\newcommand{\iceExtremeGap}{0.07}
\newcommand{\iceObs}{OSI-SAF CDR v3p0 (ICDC)}
\newcommand{\kokkosFourNodeSYPD}{80}
\newcommand{\kokkosOneNodeSYPD}{25}
\newcommand{\kokkosOneNodeStep}{0.200}
\newcommand{\kokkosTwoNodeSYPD}{46}
\newcommand{\kokkosTwoNodeStep}{0.107}
\newcommand{\meanNmonths}{360}
\newcommand{\meanWindow}{1980--2009}
\newcommand{\meshCoreDt}{1800}
\newcommand{\meshCoreLanes}{6}
\newcommand{\meshCoreLayers}{47}
\newcommand{\meshCoreResArea}{86}
\newcommand{\meshCoreResMedian}{46.6}
\newcommand{\meshCoreVertices}{127\,k}
\newcommand{\meshDarsDt}{240}
\newcommand{\meshDarsLanes}{177}
\newcommand{\meshDarsLayers}{56}
\newcommand{\meshDarsResArea}{17}
\newcommand{\meshDarsResMedian}{10.4}
\newcommand{\meshDarsVertices}{3.16M}
\newcommand{\meshFarcDt}{1200}
\newcommand{\meshFarcLanes}{30}
\newcommand{\meshFarcLayers}{47}
\newcommand{\meshFarcResArea}{85}
\newcommand{\meshFarcResMedian}{5.6}
\newcommand{\meshFarcVertices}{638\,k}
\newcommand{\meshForcaDt}{240}
\newcommand{\meshForcaLanes}{147}
\newcommand{\meshForcaLayers}{69}
\newcommand{\meshForcaResArea}{17}
\newcommand{\meshForcaResMedian}{13.9}
\newcommand{\meshForcaVertices}{2.13M}
\newcommand{\meshNgDt}{240}
\newcommand{\meshNgLanes}{511}
\newcommand{\meshNgLayers}{69}
\newcommand{\meshNgResArea}{10}
\newcommand{\meshNgResMedian}{5.7}
\newcommand{\meshNgVertices}{7.40M}
\newcommand{\meshSizeSpan}{58}
\newcommand{\meshWetFracMax}{80}
\newcommand{\meshWetFracMin}{48}
\newcommand{\ngEffSixtyFour}{86}
\newcommand{\ngSYPD}{1.50}
\newcommand{\ngStep}{0.44}
\newcommand{\ngStepOneTwentyEight}{0.35}
\newcommand{\nhIceDiffMax}{0.06}
\newcommand{\nhIceDiffMonth}{June}
\newcommand{\nhIceDiffPct}{0.5}
\newcommand{\nhMarFor}{15.9}
\newcommand{\nhMarJax}{15.9}
\newcommand{\nhMarObs}{14.1}
\newcommand{\nhRMSD}{1.62}
\newcommand{\nhSepFor}{6.9}
\newcommand{\nhSepJax}{6.9}
\newcommand{\nhSepObs}{5.3}
\newcommand{\oceanVol}{1.32}
\newcommand{\ohcDiffEnd}{+0.48}
\newcommand{\ohcDiffMax}{0.96}
\newcommand{\sProfDiffDepth}{15}
\newcommand{\sProfDiffMax}{$3.8\times10^{-4}$}
\newcommand{\sSectDiffJF}{$1.4\times10^{-4}$}
\newcommand{\sSectRMSEfor}{0.077}
\newcommand{\sSectRMSEjax}{0.077}
\newcommand{\sSectRMSEpair}{0.077 in both runs}
\newcommand{\sSectRMSEpairShort}{0.077}
\newcommand{\sbarDiffEnd}{$+2.9\times10^{-5}$}
\newcommand{\sbarDriftFor}{$+2\times10^{-5}$}
\newcommand{\sbarDriftJax}{$+5\times10^{-5}$}
\newcommand{\shIceDiffMax}{0.16}
\newcommand{\shIceDiffMonth}{November}
\newcommand{\shIceDiffPct}{1.1}
\newcommand{\shMarFor}{2.1}
\newcommand{\shMarJax}{2.1}
\newcommand{\shMarObs}{2.9}
\newcommand{\shRMSD}{0.78}
\newcommand{\shSepFor}{17.4}
\newcommand{\shSepJax}{17.3}
\newcommand{\shSepObs}{16.1}
\newcommand{\sssDiffJF}{0.002}
\newcommand{\sssRMSEfor}{0.38}
\newcommand{\sssRMSEjax}{0.38}
\newcommand{\sssRMSEpair}{0.38 in both runs}
\newcommand{\sssRMSEpairShort}{0.38}
\newcommand{\sstDiffJF}{0.004}
\newcommand{\sstRMSEfor}{0.61}
\newcommand{\sstRMSEjax}{0.61}
\newcommand{\sstRMSEpair}{0.61\,$^\circ$C in both runs}
\newcommand{\sstRMSEpairShort}{0.61\,$^\circ$C}
\newcommand{\tProfDiffDepth}{255}
\newcommand{\tProfDiffMax}{$1.3\times10^{-3}$}
\newcommand{\tSectDiffJF}{$1.2\times10^{-3}$}
\newcommand{\tSectRMSEfor}{0.379}
\newcommand{\tSectRMSEjax}{0.379}
\newcommand{\tSectRMSEpair}{0.379\,$^\circ$C in both runs}
\newcommand{\tSectRMSEpairShort}{0.379\,$^\circ$C}
\newcommand{\tbarDiff}{$+8.8\times10^{-5}$}
\newcommand{\tbarEndFor}{3.619}
\newcommand{\tbarEndJax}{3.619}
\newcommand{\tbarEndPair}{3.619\,$^\circ$C in both runs}
\newcommand{\tbarEndPairShort}{3.619\,$^\circ$C}
\newcommand{\tbarOwnDrift}{$-1.7\times10^{-2}$}
\newcommand{\tbarSevenDiffMax}{$5.4\times10^{-3}$}
\newcommand{\tbarStart}{3.636}
\newcommand{\thrHalfKeep}{79}
\newcommand{\thrPlateau}{24}
\newcommand{\thrPlateauGH}{57}
\newcommand{\thrPlateauHi}{27}
\newcommand{\thrPlateauLo}{21}
\newcommand{\thrPlateauRatio}{2.3}
\newcommand{\thrPlateauSpread}{11}
\newcommand{\thrPlateauVerts}{120\,000}
\newcommand{\thrSmallShard}{30}

\title{FESOM2-JAX v1.0: a differentiable shadow of the ocean--sea-ice model
FESOM2, cast onto GPUs}

\Author[1][nikolay.koldunov@awi.de]{Nikolay V.}{Koldunov}
\Author[1]{Sergey}{Danilov}
\Author[1]{Suvarchal}{Cheedela}
\Author[1]{Dmitry}{Sidorenko}
\Author[1]{Sebastian}{Beyer}
\Author[1]{Patrick}{Scholz}
\Author[1]{Ivan}{Kuznetsov}
\Author[1]{Jan}{Streffing}
\Author[1]{Aleksei}{Koldunov}
\Author[1]{Dmitrii}{Pantiukhin}
\Author[1]{Svetlana N.}{Loza}
\Author[1,2]{Thomas}{Jung}


\affil[1]{Alfred Wegener Institute, Helmholtz Centre for Polar and Marine
Research, Bremerhaven, Germany}
\affil[2]{Department of Physics and Electrical Engineering, University of Bremen, Bremen, Germany}

\runningtitle{FESOM2-JAX}
\runningauthor{Koldunov et al.}

\received{}
\pubdiscuss{}
\revised{}
\accepted{}
\published{}

\firstpage{1}
\maketitle

\begin{abstract}
We present FESOM2-JAX, a Python re-implementation of the Finite-volumE Sea ice--Ocean Model (FESOM2) in JAX. The model retains the unstructured-mesh, cell-vertex finite-volume formulation of the original, runs unchanged from a laptop CPU to 256 GPUs, and is end-to-end differentiable. FESOM2-JAX is a \emph{code shadow} of the Fortran model: a projection onto the Python ecosystem, translated with large language models and verified kernel by kernel against the original. It is built to lower the barrier to experimentation, from new numerics and parameterizations to gradient-based calibration and hybrid physics--machine-learning components, while remaining close enough to the original so that what is developed in the shadow can be transferred back. In a \hindcastYears\ hindcast at 1$^{\circ}$ equivalent resolution with identical physics and forcing, the mean states of the JAX and Fortran versions differ from each other by two orders of magnitude less than either differs from observations, and the two runs agree for six decades in global temperature, salinity, heat content, and sea ice. The complete 1$^{\circ}$ configuration fits on a single GPU, a node of four GH200 superchips integrates $\sim$\ghSypdCoreFourRound\ simulated years per wall-clock day, and meshes of up to 7.4 million surface vertices ($\sim$5\,km) scale to 128 GPUs. What limits the model is communication rather than arithmetic. What the shadow adds to the original is the gradient: a single reverse-mode pass through the full time loop returns the sensitivity of a model diagnostic to a parameter at every mesh vertex, verified against finite differences. To our knowledge, FESOM2-JAX is the first global ocean--sea-ice model of CMIP-class complexity written natively in a differentiable framework, and the first on an unstructured mesh.
\end{abstract}

\introduction
\label{sec:intro}

The Finite-volumE Sea ice--Ocean Model (FESOM2; \citealp{danilov2017fesom2}) is a global ocean--sea-ice model built on an unstructured triangular mesh. Its cell-vertex finite-volume discretization lets the horizontal resolution vary smoothly over the
globe, so a single configuration can resolve narrow straits, boundary currents, or a region of interest at eddying resolution while keeping the open ocean coarse. The model has been assessed in depth against observations and against its predecessor
\citep{scholz2019fesom2,scholz2022fesom2}, serves as the ocean component of the AWI climate model in upcoming CMIP7 \citep{streffing2022awi,moon2025earth}, part of the several kilometer-scale resolution modeling initiatives, including {nextGEMS} \citep{segura2025nextgems} and Destination Earth Climate Digital Twin \citep{destine,wedi2025}. It scales efficiently to thousands of cores in its Fortran--MPI implementation \citep{koldunov2019fesom2scaling}.

Models of this class face two slow-moving but compounding pressures. The first is hardware and its toolchain. The computational capacity of current high-performance systems (HPC) is concentrated in GPUs, and accelerator toolchains for Fortran have matured later and less uniformly than their C++ and Python counterparts; there are several community's routes around this, such as compiler directives \citep[e.g.\ ICON;][]{giorgetta2022icon}, domain-specific languages \citep{dahm2023pace,adams2019lfric}, C++ performance-portability frameworks, including our own translation of FESOM2 to C++/Kokkos \citep{koldunov2026llmport}, or new codes written GPU-native from the start \citep{ramadhan2020oceananigans,silvestri2025oceananigans}. The second pressure is people. The generation of scientists now entering the field is trained in Python and its machine-learning ecosystem. Much of model development is exploratory: trying a new advection scheme, a new parameterization, or a new calibration strategy. For such work, Python's
interactivity and libraries provide a better medium than Fortran production code, provided that the Python model remains faithful enough to the original for the results to carry over.

Our response to these pressures, begun in \citet{koldunov2026llmport} with an LLM-assisted translation of FESOM2 from Fortran through C to C++/Kokkos, is to develop and maintain \emph{code shadows}. They are defined as projections of one model onto different technology stacks (C++/Kokkos for performance portability, Python for experimentation), but kept faithful to the original that casts them rather than allowed to drift into independent models. The Fortran production model remains
the single source of truth (the ideal object, in the Platonic sense) so a
shadow is neither a fork nor a successor; it is verified by running alongside the original on identical inputs and comparing outputs. This way developments made in the shadow, such as tuned parameter sets, trained machine-learning components, or validated numerical variants, can be carried back into the Fortran production model. This paper presents the Python shadow, FESOM2-JAX: a re-implementation of FESOM2 in python JAX \citep{bradbury2018jax}. It is a NumPy-style codebase that a new researcher or student can read, run on a laptop or, unchanged, on 256 GPUs, and modify. This makes it a useful vehicle for teaching and a platform for experiments that are difficult to carry out in a Fortran production code.

The clearest example of this category is the adjoint. Adjoint models, reverse-mode derivatives of the discretized dynamics \citep{thacker1988fitting,errico1997adjoint}, they underpin, for example, the ECCO ocean state estimates based on MITgcm ocean model \citep{stammer2002ecco,wunsch2007practical,forget2015ecco}, variational data assimilation for NEMO and ROMS \citep{vidard2015nemotam,moore2011roms}, and adjoint sensitivity studies through ocean--sea-ice dynamics \citep{heimbach2010seaice}. Obtaining an adjoint from a Fortran model has, however, been challenging. It requires source-to-source transformation tools with tool-specific coding restrictions, hand-tuned checkpointing of the reverse sweep, and regeneration and re-verification whenever the forward model evolves \citep{giering1998tamc,heimbach2005adjoint,utke2008openad,griewank2000revolve}. As a result, most ocean models, including FESOM2, never acquired one, and parameter estimation has relied instead mostly on gradient-free methods \citep{sumata2019seaice,iglesias2013eki,souza2020uq}.

In a differentiable framework, by contrast, the adjoint is derived by the framework from the same source that defines the forward model \citep{baydin2018autodiff,paszke2019pytorch,bezanson2017julia}. It is exact for the discretization, remains consistent with the code as the code changes, and extends to whatever is added to the model, including neural-network components. This is what makes hybrid physics--machine-learning models trainable through the dynamics \citep{rasp2018deep,kochkov2024neuralgcm}. Differentiable programming is therefore increasingly argued to be a foundation for the next generation of Earth-system models \citep{gelbrecht2023diffprog,shen2023diffmodel}. This can be pursued either by rewriting models in differentiable frameworks, as we do here, or by differentiating existing compiled code through the compiler, the route pursued, for example, with Enzyme by the DJ4Earth project \citep{moses2020enzyme,moses2026dj4earth}.

FESOM2-JAX belongs to a growing family of geophysical models written natively in Python and JAX. Veros \citep{hafner2018veros} translated the pyOM2 primitive-equation model into Python and, with a JAX backend, runs global structured-grid configurations on GPUs at throughput competitive with its Fortran ancestor, including multi-node runs communicating through MPI \citep{hafner2021veros,hafner2021mpi4jax}. Veris \citep{gartner2026veris} has recently added sea ice to the same ecosystem by re-implementing the MITgcm sea-ice model in JAX, matching hundreds of CPU cores on a single GPU. The translation of legacy Fortran into this ecosystem is also becoming more systematic. Large-language-model-assisted conversions to JAX have been demonstrated for a CESM photosynthesis module \citep{zhou2024fortranjax} and, more systematically, for a 19,000-line land-surface model \citep{lahlou2026lsmjax}, with both gaining GPU execution and access to gradients. This is the same class of tooling with which the code-shadow lineage behind the present work was built \citep{koldunov2026llmport}. Differentiable solvers of reduced complexity, including quasi-geostrophic models \citep{yan2025qgonline} and engineering CFD \citep{kochkov2021mlcfd,bezgin2023jaxfluids}, have become testbeds for subgrid closures trained ``online'', through the solver, which outperform the same closures fitted offline \citep{frezat2022posteriori}. In the atmosphere, NeuralGCM \citep{kochkov2024neuralgcm} couples a differentiable spectral dynamical core in JAX to a learned physics package and trains the hybrid model end to end against reanalysis. Within this family, FESOM2-JAX is, to our knowledge, the first global ocean--sea-ice model of CMIP-class complexity: the full parameterization suite of such an ocean model and its dynamic--thermodynamic sea ice in a single differentiable time loop. It is also the first built on an unstructured mesh.

The body of the paper is organized around the three requirements that FESOM2-JAX must meet for a code shadow to be useful: reproducible experimentation, scientific fidelity, and computational performance.

\begin{enumerate}
\item \textbf{Reproducible experimentation} (Sect.~\ref{sec:usability}). A simulation is specified by a fully declarative YAML-based run configuration. The model supports restarts and basic diagnostics and provides all the functionality needed to conduct multi-decadal or high-resolution integrations within conventional Earth-system model data workflows. A continuous-integration ensures that the default configuration remains bit-reproducible across releases.
\item \textbf{Scientific fidelity} (Sect.~\ref{sec:fidelity}). In a
\hindcastYears\ hindcast on a mesh with an equivalent resolution of 1$^{\circ}$, configured identically to a Fortran run, the JAX model closely reproduces the Fortran climatology. The comparison is statistical and includes climatological bias maps, multi-decadal drift, and sea-ice metrics.
\item \textbf{Computational performance} (Sect.~\ref{sec:performance}). The model runs on meshes ranging from $\sim10^5$ to $\sim7\times10^6$ surface vertices and on hardware ranging from a laptop CPU to 256 GPUs. It is fast in absolute terms but remains communication-bound.
\end{enumerate}

Section~\ref{sec:model} first describes the model, its numerics, and the procedure by
which the port was verified; Sects.~\ref{sec:usability}--\ref{sec:performance}
then address the three requirements in turn. Section~\ref{sec:outlook} demonstrates
the differentiability of the full time loop. The applications enabled by this capability,
including gradient-based calibration, hybrid physics--machine-learning modelling, and
the transfer of their products back into the Fortran original, are the subject of
ongoing work and lie beyond the scope of this model-description paper.

\section{Model and numerics}
\label{sec:model}

FESOM2-JAX solves the same equations as the Fortran version of FESOM2, on the same geometry, with the same discretization and the same production physics. The two codes therefore differ in implementation, not in formulation, and everything described below is common to both unless explicitly noted. The port covers the components required for the forced ocean--sea-ice configuration of FESOM2; the components that are not ported are listed in Sect.~\ref{sec:scope}. Table~\ref{tab:components} summarizes the ported components and the settings used in this study, and Appendix~\ref{app:components} describes them one by one. The port itself was carried out kernel by kernel, following the LLM-agentic workflow described by \citet{koldunov2026llmport}; in the present work, the harness was Claude Code and the language models used were Opus 4.8 and Fable.


\subsection{FESOM2 and the scope of the port}
\label{sec:scope}

FESOM2 solves the hydrostatic, Boussinesq primitive equations on an unstructured
triangular mesh with a cell-vertex finite-volume discretization. Scalar fields,
including temperature, salinity, pressure, and sea-surface height, are located at mesh
vertices and advanced on the control volumes, while horizontal velocity is located at
triangle centroids. The vertical coordinate is the arbitrary Lagrangian--Eulerian (ALE)
coordinate, used throughout this study in its $z^\star$ form
\citep{adcroft2004rescaled}. Sea ice is a single-class dynamic--thermodynamic model,
based on the Finite-Element Sea Ice Model \citep[FESIM;][]{danilov2015fesim} and
discretized at the same mesh vertices as the ocean scalars. The discretization and numerics are
described by \citet{danilov2017fesom2}, and further model elements, together with their
assessment against observations and against the predecessor model, by
\citet{scholz2019fesom2,scholz2022fesom2}. We do not reproduce that material here. Appendix~\ref{app:components} gives the
component-by-component description for readers who want it without consulting those
papers.

The scope of the present model is the forced ocean--sea-ice problem. FESOM2-JAX is
driven by prescribed atmospheric forcing and has no tides, icebergs, ocean
biogeochemistry, ice-shelf cavities, floating ice, or partial bottom cells. Among the components that are
ported, several of FESOM2's alternatives are retained and are chosen from the
configuration file: the ALE coordinate is ported in its $z^\star$ form, with the linear
free surface as the alternative surface treatment, while FESOM2's z-level mode is not
ported; vertical mixing can be run with the
turbulent-kinetic-energy closure of \citet{gaspar1990tke} as implemented in the CVMix
library \citep[cvmix-TKE;][]{griffies2015cvmix,vanroekel2018kpp}, with the K-profile parameterization
\citep{large1994kpp}, or with the Richardson-number scheme of
\citet{pacanowski1981pp}; and the sea-ice rheology can be either mEVP or standard EVP. Atmospheric forcing is the one component in which the port is
narrower than the original, in that the JRA55-do reader \citep{tsujino2018jra55do} is
its only forcing implementation. These choices define the boundaries of a first
model-description paper, not limitations of the approach.

\subsection{The JAX implementation}
\label{sec:jax-implementation}

The implementation is ordinary python JAX array code \citep{bradbury2018jax}, using JAX
version 10.1. Three choices shape it: the timestep is a pure function of the model state
and compiles to a single program, differentiability is treated as a property of every
kernel rather than as a later addition, and one source runs unchanged on any device
count.

\paragraph{One pure step, one compiled program.} The model state, including every prognostic
field, the Adams--Bashforth history, and the warm-start vector of the CG solver, is
stored as a single immutable pytree, and one timestep is a pure function of that state.
Fields are stored as dense vertex-by-level arrays. The number of wet layers varies from
column to column with the bottom topography, and that variation is carried by masking
the cells below the bottom rather than by storing only the wet ones, so every array has
a fixed shape and the XLA (Accelerated Linear Algebra) compiler can fuse the full
timestep into one compiled program. The time loop is a single
\texttt{jax.lax.scan} over this function, compiled once and iterated without returning
to the Python interpreter. A long campaign is then a chain of such scans, reusing the
compiled executable across job boundaries. Configuration is static at compile time:
an inactive scheme, for example KPP when TKE is selected, is absent from the compiled
program rather than branched over at run time. An all-on run therefore incurs no cost
for options it does not use. All state and arithmetic are in float64, matching the
Fortran model; this is required both for multi-decadal conservation and for the
verification procedure of Sect.~\ref{sec:verification}.

\paragraph{Differentiability by construction.} Reverse-mode differentiability is
treated as a correctness property of every kernel, rather than as a feature added
afterwards. Three patterns recur. First, iterative solvers with data-dependent stopping
are avoided in favour of fixed iteration counts: five passes of the bulk-flux stability
iteration, five Newton steps in the ice thermodynamics, and 120 mEVP iterations. This
makes the backward sweep a well-defined unroll. The one exception is the CG solver for
the free surface, which is wrapped in JAX's \texttt{custom\_linear\_solve} so that the
forward pass reproduces the reference's early-stopped iterate, while the backward pass
differentiates the linear system implicitly instead of unrolling the iteration.
Second, non-smooth primitives are given finite gradients at their switching points:
guarded square roots and powers at zero, clamped denominators, and identity rows in
the tridiagonal solves for dry cells so that no gradient leaks below the bottom.
Third, the memory of the reverse sweep over long integrations is contained by nested
\texttt{jax.checkpoint} blocks around groups of timesteps
(Sect.~\ref{sec:outlook}).

\paragraph{One source, any device count.} Distributed execution uses
\texttt{jax.shard\_map} over a one-dimensional device axis. The domain decomposition is
FESOM2's own: each device holds the entities it owns, followed by a halo rim of entities
it reads but does not own, padded to a common size so that the per-device arrays keep a
fixed shape. The layout is the familiar overlap arrangement of a Fortran ocean model. The
difference lies in the exchange itself, which is not a call into a message-passing library
but an operation written in the same language as the rest of the model and compiled
together with it. We implemented four transports, described in Sect.~\ref{sec:halo}.
Global sums are \texttt{psum} reductions.

The same source also defines the distributed adjoint. In a Fortran model the adjoint of
the exchange is written and maintained by hand. This holds whether the surrounding adjoint
is generated by source transformation, as for the MIT general circulation model
\citep{heimbach2005adjoint}, in which case the exchange is a side-effecting subroutine
wrapping message-passing calls that the tool cannot traverse, or hand-coded throughout, as
for ROMS and NEMO \citep{moore2011roms,vidard2015nemotam}; the handling of parallelisation
is among the reasons given for the latter choice \citep{vidard2015nemotam}. The underlying
mathematics is the same in every case: an exchange is linear in the field it moves, so its
adjoint is its transpose, in which the adjoint values of the halo copies are returned to
the owner and summed there. In the present case the exchange is free of side effects and
is built from primitives whose transposes the framework already defines, so that this
transpose is obtained by composition rather than by hand. This reduces the amount that
must be verified but does not by itself establish correctness. We verify the distributed
adjoint by requiring the sharded gradient to match the single-device gradient, which is
itself checked against finite differences (Sect.~\ref{sec:verification}). The broadcast,
padded, and coloured transports carry that adjoint; the ragged all-to-all is used
forward-only, because we found its reverse-mode rule to be defective in JAX 10.1.

The step body is unchanged between a laptop CPU, a single GPU, and 256 GPUs. Running the
sharded code path on one device reproduces the single-device path bit for bit, and this
identity is the invariant by which we verify the distributed implementation. Restarts and
output are written in the mesh's global vertex numbering rather than in any device's local
order, so that the files a run produces do not depend on the number of devices used: a
restart written on four GPUs can be read on 256, and output from either can be compared
directly (Sect.~\ref{sec:usability}).

\subsection{The halo exchange, four ways}
\label{sec:halo}
\begin{figure*}[t]
\includegraphics[width=\textwidth]{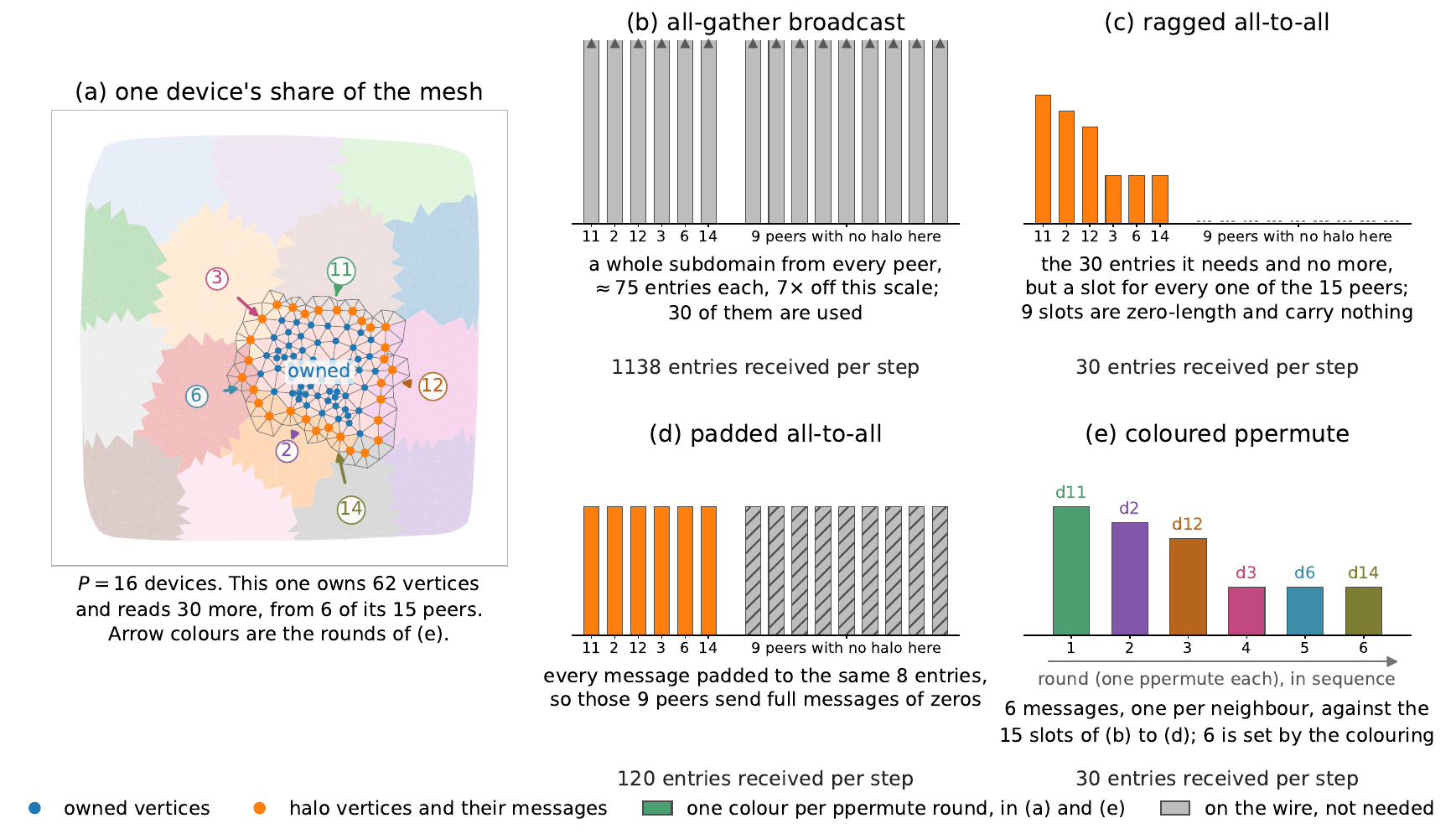}
\caption{What each halo transport sends, for one device of a schematic partition.
(a) A small unstructured mesh cut into $P=16$ subdomains. The highlighted device owns
62 vertices (blue) and its stencils read 30 more that it does not own (orange): its
halo. Six of the other 15 devices own part of that halo and are numbered; the remaining
nine own none of it. (b to d) What the highlighted device receives per step under each
of the three exchanges, one bar per device, all panels on the same vertical scale. The
all-gather delivers every other device's whole subdomain, which runs off the top of the
scale, and 30 of the 1138 values received are used. The ragged all-to-all delivers
exactly the halo entries each device owns and nothing more, but still carries a slot for
all 15; the nine drawn as dashes are empty and move no data. This is the transport with
no usable adjoint. The padded all-to-all delivers the same entries but requires every
slot to be the same size, so those nine devices send full-sized messages of zeros that
do cross the wire. (e) The coloured \texttt{ppermute} is a schedule rather than a single
exchange, so its bars are rounds rather than devices: the halo arrives in $K$ rounds,
and in each round every device exchanges with exactly one partner. The arrows in (a) are
coloured by round, and each bar is labelled with the partner it serves. It sends six
messages against fifteen slots, and $K$ is set by the number of neighbours rather than
by the device count, which is what keeps its cost flat as the partition is spread wider;
but the rounds run one after another, so it pays $K$ start-up costs where a single
exchange pays one. The totals are exact for this partition; on the meshes of
Sect.~\ref{sec:meshes} a device has 6 to 14 neighbours while $P$ reaches 256, which
widens the same gaps. Table~\ref{tab:halo} gives the measured per-step times.}
\label{fig:halo}
\end{figure*}

Because the halo exchange lies on the critical path of every stencil, we implemented it
four ways; Fig.~\ref{fig:halo} shows what each of them puts on the wire. They differ
along two axes. The
first is whether the adjoint is exact. The all-gather broadcast, the padded all-to-all,
and the coloured \texttt{ppermute} all differentiate correctly, because the adjoint of a
halo read is an additive scatter back to the owner. By contrast,
\texttt{lax.ragged\_all\_to\_all}, which ships the least data, has a defective
reverse-mode rule in JAX 10.1 and is usable forward-only. The second axis is how the
per-device wire volume grows with the device count $P$. The all-gather moves the whole
field, however little of it is halo. The ragged and padded all-to-alls carry one message
slot for every device, whether or not that device owns any of the halo, so their cost
rises as the partition is spread over more of them; the padded variant also pads every
slot to a common size, so the devices that own none of the halo send full-sized messages
of zeros. The coloured transport instead exchanges with one partner at a time, in $K$
rounds coloured so that no device is asked to talk to two partners at once, and its cost
is bounded by the largest number of neighbours any device has, 6 to 14 on the meshes
used here. That bound does not grow with $P$, because a spatial partition keeps a
bounded number of neighbours at any scale.
All but the ragged transport also run on the CPU backend. Section~\ref{sec:perf-halo}
measures the four against each other.

\subsection{Verification of the port}
\label{sec:verification}

A port of this size is only as credible as its verification, so we describe the
verification procedure and the reference from which the port was made. The proximate
reference for FESOM2-JAX was not the Fortran source directly, but the instrumented
serial C translation of FESOM2 produced as the intermediate of the Fortran-to-Kokkos
port of \citet{koldunov2026llmport}: a kernel-by-kernel mirror of the Fortran code
that writes every intermediate field of every substep to disk. For the present work,
that C reference was extended beyond the configuration of the Kokkos study with the
components used in this paper: the $z^\star$ coordinate, the cvmix-TKE closure, and
the mEVP sea-ice rheology. Each extension was itself a literal translation of the
corresponding Fortran routine, validated against instrumented Fortran output before
being adopted as a baseline. The unit of porting was then the kernel, corresponding to
one Fortran subroutine's worth of physics, and every JAX kernel was gated against the
C reference's dumps. JAX does not reproduce the loop order of the reference, so
agreement cannot be exact, and the tolerance is therefore set by what the kernel does.
Kernels that work point by point, or that only read values at neighbouring points, are
required to agree to about $10^{-15}$ relative, which is the level of rounding alone.
Kernels that accumulate contributions, such as the assembly of edge fluxes onto the
control volumes and the global sums, are added up in a different order by JAX and are
required to agree to about $10^{-12}$.
Multi-step replay tests then verify the state threading that single-step tests cannot
see, including the Adams--Bashforth history, the CG warm start, and the
layer-thickness commit, and the sharded code path must reproduce the dense path
exactly on one device and to reassociation tolerance on many.

Differentiability is
tested with the same rigour as the forward model: gradients are checked against finite
differences kernel by kernel, at states away from the switching points at which the
model is not differentiable, and through
short full-model integrations. The suite comprises 78 test modules (roughly 640
tests), of which 41 modules exercise gradients; the continuous-integration gate of
Sect.~\ref{sec:usability} additionally pins the shipped configuration to a stored
reference trajectory bit for bit. Finally, the chain closes back on the original: the
hindcast of Sect.~\ref{sec:fidelity} compares the assembled model against Fortran
FESOM2 itself, rather than against the C reference the port was built from, so an
infidelity introduced at either translation step, Fortran to C or C to JAX, would
appear in this comparison. The same procedure localized the freshwater-budget defect
of Sect.~\ref{sec:fidelity}. Because every kernel had already been verified on its own,
a drift that appears only over climate timescales cannot originate inside a kernel and
must instead arise where two kernels are joined. Checking which of the conservation
budgets fails to close then locates the defect.

\subsection{Meshes}
\label{sec:meshes}

Five meshes are used in this paper, and they are the same community FESOM2 meshes the
Fortran model runs on, each introduced in the study cited beside it in
Table~\ref{tab:meshes}; the geometry tabulated there is measured from the mesh files the
model reads. They span two orders of magnitude in size, from a $1^\circ$-equivalent
configuration to a mesh with 7.4 million surface vertices, and they follow different
design principles. FORCA20 is a Mercator mesh, on which resolution is set by latitude
and increases gradually towards the poles. DARS follows the local Rossby radius
\citep{sein2017rossby} and the energetically active regions. NG5 is close to uniform at
5\,km over most of the globe and is coarsened only where the flow is less energetic.
fArc is CORE2 refined in a single region, the 4.5\,km Arctic.

One property of the hierarchy matters for the rest of the paper. The vertical extent
grows with the horizontal one, from 47 layers on CORE2 to 69 on NG5, so the cost of a
mesh grows faster than its surface-vertex count: NG5 has \meshSizeSpan\ times CORE2's
vertices but 86 times its vertex-levels, the product of vertices and layers that the
model actually computes on. CORE2 is the mesh of the hindcast in
Sect.~\ref{sec:fidelity} and the only mesh on which the GM/Redi parameterization is
active in this study; FORCA20 carries the eddying comparison of Sect.~\ref{sec:fidelity};
and all five are used in the performance measurements of Sect.~\ref{sec:performance}.
Appendix~\ref{app:meshes} notes how far the nominal resolution of
Table~\ref{tab:meshes} departs from the area-weighted one.

\begin{table*}[t]
\caption{Model components ported in FESOM2-JAX and the schemes used in this study.
The Gent--McWilliams/Redi parameterization is active in the CORE2 hindcast and switched
off in the configurations used for the performance measurements; all other components
are common to every configuration.}
\label{tab:components}
\small
\begin{tabular}{l p{0.40\textwidth} p{0.30\textwidth}}
\tophline
Component & Scheme / setting & Reference \\
\middlehline
Vertical coordinate  & ALE, $z^\star$ (linear free surface also ported) & \citet{adcroft2004rescaled} \\
Equation of state    & polynomial, in-situ density & \citet{jackett1995eos} \\
Pressure gradient    & density-Jacobian & \citet{shchepetkin2003pgf} \\
Free surface         & semi-implicit ($\alpha=\theta=1$), preconditioned CG to $10^{-5}$, warm-started & \citet{danilov2017fesom2,marshall1997mitgcm} \\
Time stepping        & Adams--Bashforth 2 ($\epsilon=0.1$); implicit vertical mixing & \\
Tracer advection     & MUSCL-type--FCT; 3rd-order horizontal, 4th-order vertical & \citet{danilov2017fesom2,zalesak1979fct} \\
Vertical mixing      & cvmix-TKE, $c_k=0.1$, $c_\epsilon=0.7$ (KPP, PP also ported) & \citet{gaspar1990tke,griffies2015cvmix,vanroekel2018kpp} \\
Mesoscale eddies     & GM (streamfunction BVP) $+$ Redi, coarse mesh only & \citet{gent1990isopycnal,ferrari2010bvp} \\
Horizontal viscosity & biharmonic filter, flow-aware coefficient & \citet{danilov2017fesom2,juricke2020kinematic} \\
Sea-ice dynamics     & mEVP, $\alpha=\beta=250$, 120 iterations (EVP also ported) & \citet{bouillon2013mevp,kimmritz2015mevp} \\
Sea-ice thermodynamics & zero-layer, 7 thickness classes & \citet{semtner1976thermo,hibler1979seaice} \\
Sea-ice advection    & Taylor--Galerkin FE-FCT & \citet{lohner1987fct} \\
Atmospheric forcing  & NCAR bulk formulae, JRA55-do & \citet{large2009bulk,tsujino2018jra55do} \\
Surface salinity     & relaxation to monthly climatology, 10\,m per 60 days & \\
Precision            & float64 throughout & \\
Time integration     & single \texttt{jax.lax.scan}; \texttt{shard\_map} over devices & \\
\bottomhline
\end{tabular}
\belowtable{}
\end{table*}

\begin{table*}[t]
\caption{The meshes used in this study. \emph{Layers} is the number of vertical layers and \emph{vertex-levels} the product of the two, which is the number of entries in the model's dense state arrays: because the varying number of wet layers is carried as masks rather than by storing only the wet cells (Sect.~\ref{sec:jax-implementation}), every timestep advances all of those entries, and not only the \meshWetFracMin--\meshWetFracMax\,\% of them that are wet ocean, so the full product is the work a timestep performs. \emph{Resolution} is the nominal resolution each mesh is designed and named for, with the region it applies to where the mesh is regionally refined; the vertex, layer and vertex-level counts are measured from the mesh files the model reads. $\Delta t$ is the production timestep, at which all throughput in Sect.~\ref{sec:performance} is reported. Measured per-step cost on each mesh is given in Table~\ref{tab:perf}.}
\label{tab:meshes}
\small
\begin{tabular}{lrrrlrl}
\tophline
 & surface & & vertex-levels & & $\Delta t$ & \\
Mesh & vertices & layers & ($10^6$) & resolution & (s) & Reference \\
\middlehline
CORE2 & 126\,858 & 47 & 6.0 & 1$^\circ$ equivalent & 1800 & \citet{wang2014fesom} \\
fArc & 638\,387 & 47 & 30.0 & 4.5\,km, Arctic & 1200 & \citet{wang2018arctic} \\
FORCA20 & 2\,127\,871 & 69 & 146.8 & 1/5$^\circ$ equivalent & 240 & \citet{koldunov2026lec} \\
DARS & 3\,160\,340 & 56 & 177.0 & $\sim$10\,km & 240 & \citet{streffing2022awi} \\
NG5 & 7\,402\,886 & 69 & 510.8 & $\sim$5\,km & 240 & \citet{rackow2025ifsfesom} \\
\bottomhline
\end{tabular}
\belowtable{}
\end{table*}

\section{Usability and reproducibility}
\label{sec:usability}

For the Python JAX shadow, usability is a design requirement rather than a byproduct. This section describes the design choices intended to make FESOM2-JAX a practical tool rather than a research prototype: single-file configuration, restarts, data output, and an explicitly scoped reproducibility guarantee. In the resulting workflow, a single YAML
\texttt{RunConfig} drives one invocation: load a restart or an initial condition, integrate $N$ steps as a \texttt{jax.lax.scan}, stream online per-period means and subsampled snapshots, and write restarts.
A scheduler dependency then chains these invocations across batch jobs to assemble a long integration.

\subsection{One run, one file}

A simulation is specified by a single YAML \texttt{RunConfig}: the mesh and its partition, the physics, the timestep, the forcing, and the output and restart targets. Physics components are enabled by their presence and configured by their contents. Mutually exclusive choices, such as the two vertical-mixing schemes, are rejected at load time rather than at run time. Every key is optional, and an absent key falls back to a default. This property is enforced by a regression test, so that adding configurability does not silently change an existing result. The abridged listing below is the all-on CORE2 configuration used for the hindcast of
Sect.~\ref{sec:fidelity}:

\begin{verbatim}
ale: {}                  # z* vertical coordinate (absent => linear free surface)
tke: {}                  # cvmix-TKE vertical mixing (exclusive with kpp)
ice: {whichEVP: 1}       # sea ice, modified-EVP rheology
gm:  {}                  # Gent-McWilliams + Redi (coarse mesh)
visc: {}                 # flow-aware viscosity, default coefficients
tracer: {}               # FCT tracer advection
dt: 1800.0               # timestep [s]
mesh: data/mesh_core2
partition: dist_4        # device count of the on-disk partition
forcing: {kind: core2, start_year: 1958}   # JRA55-do, real calendar
snapshot_every: 48       # steps (~daily); 0 = off
checkpoint_every: 1440   # rolling-restart cadence (~monthly)
restart_in:  null        # null => cold start from climatology
restart_out: runs/core2/restart
restart_archive_out: runs/core2/restart_archive
restart_archive_period: year   # immutable archival restarts
duration: 10yr           # or an explicit step count
\end{verbatim}

One invocation performs a single task: it loads an initial condition or restart, integrates for a requested number of steps or a specified duration (``\texttt{10yr}'', ``\texttt{3mo}''), streams diagnostics, writes restarts, and exits. Multi-job integrations, such as the 62-year hindcast below, are assembled outside the model by an ordinary batch-scheduler dependency chain, so no bespoke driver is required.

\subsection{Restarts and output}

The model writes two restart streams for different purposes. A \emph{rolling} restart,
written every \texttt{checkpoint\_every} steps into a single directory, supports crash
recovery and the handoff between chained batch jobs, while an \emph{archival} stream
writes an immutable restart at calendar boundaries, yearly here and monthly for the
high-resolution runs, under a unique name, \texttt{fesom.<YYYY>.<DDD>.<SSSSS>}, together
with a \texttt{restart.latest} pointer. Diagnostics are computed online as
per-calendar-period means, with daily and monthly streams whose field lists are
configurable, and are written alongside time-subsampled snapshots in a single pass and
with constant memory use. Restarts and output alike are written in the mesh's global
vertex numbering rather than in any device's local order, so that a run checkpointed on
$N$ GPUs can be resumed on any number $M$, and a field can be analysed without knowing
the decomposition it was produced on. Output is written as plain Zarr stores with vertex
coordinates and a CF-convention time axis on a real calendar, so that a store can be
opened directly in xarray or in an unstructured-mesh viewer.

\subsection{Reproducibility and its scope}

The shipped baseline configuration reproduces a stored reference trajectory bit for bit, and a continuous-integration test fails on any deviation. This gate guarantees that a given release computes a specific, reproducible result, and it is one reason the model is kept in float64 throughout. The guarantee holds on a single device. Multi-GPU runs are reproducible only to round-off, because the reduction order of the collectives is not associative
(Sect.~\ref{sec:fidelity} accounts for this in the comparison methodology). The model, a user guide, and the tutorial notebook are archived as a tagged release (see Code availability).

\section{Fidelity: a CORE2 \hindcastYears\ hindcast against Fortran FESOM2}
\label{sec:fidelity}


\subsection{Experimental design}
\label{sec:design}

We run FESOM2-JAX and Fortran FESOM2 in the same configuration: the CORE2 mesh
(Table~\ref{tab:meshes}), \hindcastYears\ forced by the JRA55-do reanalysis
\citep{tsujino2018jra55do}, a cold start from the PHC3.0 winter climatology
\citep{steele2001phc}, the $z^\star$ coordinate with cvmix-TKE vertical mixing, mEVP
sea ice and GM/Redi (Table~\ref{tab:components}), and $\Delta t = 1800$\,s.
The comparison is against Fortran FESOM2, not against the
C reference from which the port was built (Sect.~\ref{sec:verification}). Both translation steps are therefore under test: an
error introduced in the Fortran-to-C step would be invisible in a comparison against C,
but appears here.

The ocean is chaotic and the two runs are not bit-identical, so their internal
variability diverges however faithful the port is (Sect.~\ref{sec:usability}). We therefore compare
statistics rather than instantaneous fields: the time-mean state, the biases against
observations, the multi-decadal drift and the sea-ice cycle. Monthly-mean climatologies
are compared over \meanWindow\ and integrals over the full hindcast.

\subsection{Mean state}
\label{sec:mean}

\begin{figure*}[t]
\includegraphics[width=\textwidth]{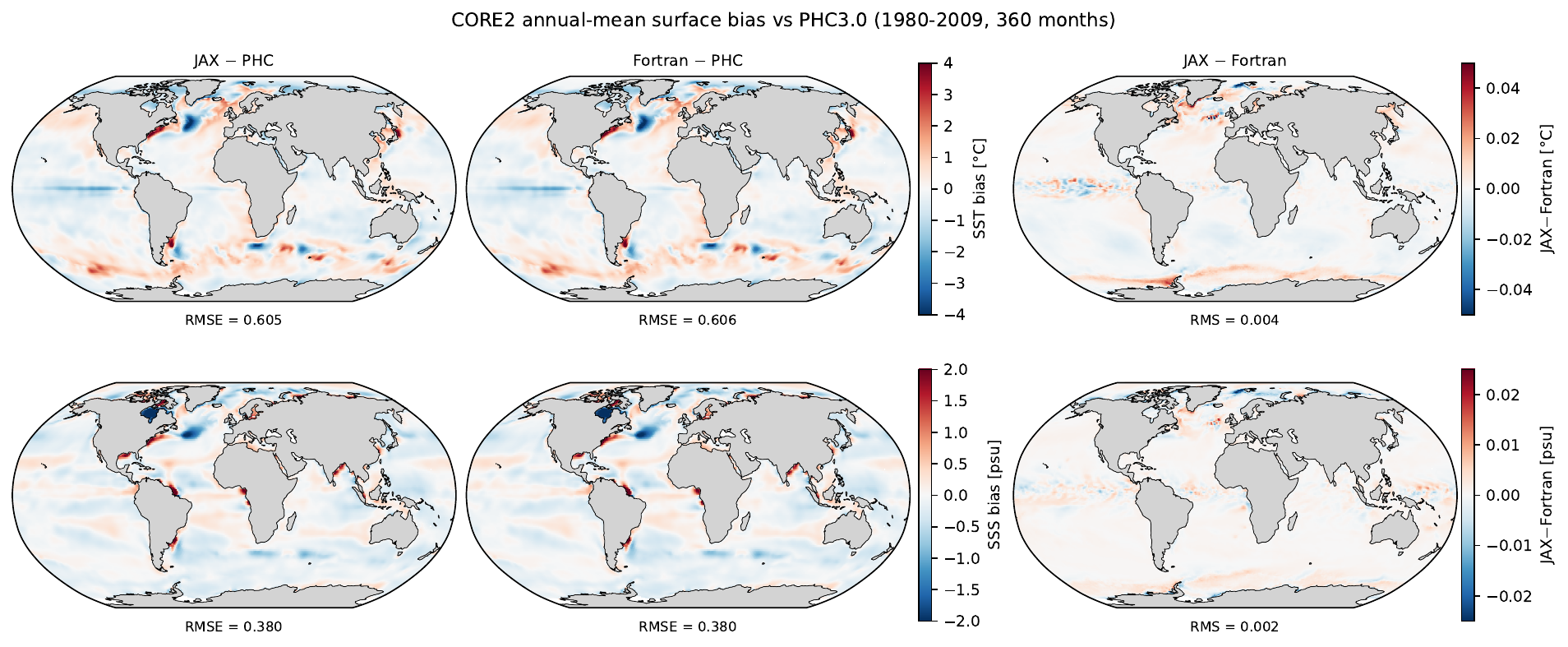}
\caption{CORE2 \meanWindow\ annual-mean surface bias against PHC3.0. Rows: SST (top),
SSS (bottom). Columns: JAX$-$PHC, Fortran$-$PHC, and JAX$-$Fortran. Note the much
tighter colour bar for the difference column ($\pm0.05\,^\circ$C, $\pm0.025$). RMSE is
annotated.}
\label{fig:meanstate}
\end{figure*}

Figure~\ref{fig:meanstate} shows the \meanWindow\ annual-mean surface bias of each model
against PHC3.0. The two bias patterns are visually indistinguishable, with
root-mean-square (RMS) errors against observations of \sstRMSEpair\ for sea-surface
temperature (SST) and \sssRMSEpairShort\ for salinity (SSS). Both models carry the familiar
CORE2 mean-state errors: the warm Gulf Stream-separation bias, the cold subpolar gyres,
and the Southern Ocean pattern.

The JAX$-$Fortran difference in the third column is two orders of magnitude below either
model's bias against observations, at \sstDiffJF\,$^\circ$C and \sssDiffJF\ RMS.
It is not spatially uniform but concentrated along the western
boundary currents, the Southern Ocean fronts and the marginal ice zones. That structure
may be worth investigating, but at this amplitude it has no practical consequence.

\begin{figure*}[t]
\includegraphics[width=\textwidth]{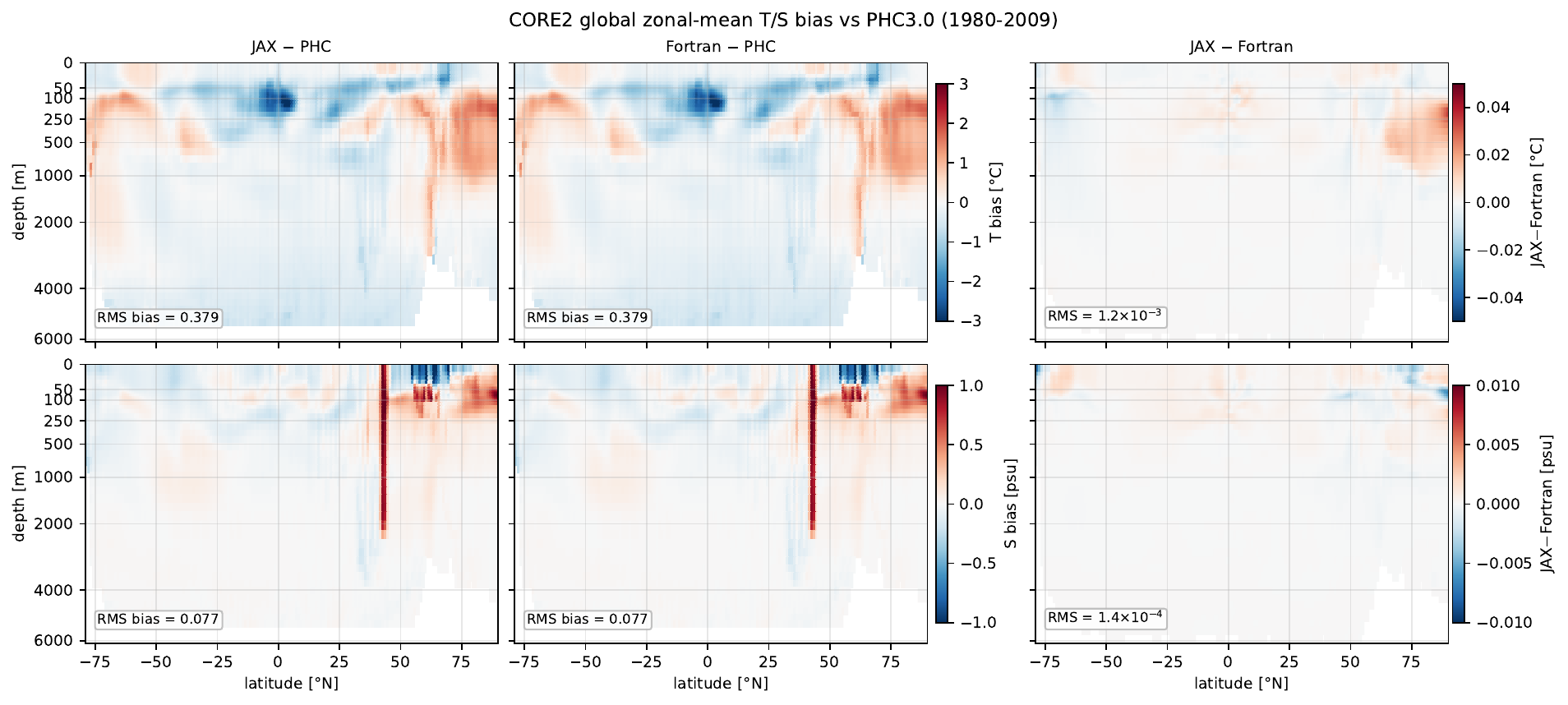}
\caption{CORE2 global zonal-mean temperature (top) and salinity (bottom) bias against
PHC3.0 over \meanWindow, on a sqrt-warped depth axis. Columns: JAX$-$PHC,
Fortran$-$PHC, and JAX$-$Fortran. Note the much tighter colour bar for the difference
column ($\pm0.05\,^\circ$C, $\pm0.01$). The annotated RMS is that of the zonal-mean
bias, so it is smaller than a full three-dimensional RMS.}
\label{fig:sections}
\end{figure*}

Figure~\ref{fig:sections} repeats the comparison in the vertical, as global zonal-mean
sections. The bias sections are again indistinguishable, with RMS biases of
\tSectRMSEpair\ for temperature and \sSectRMSEpairShort\ for salinity. The
JAX$-$Fortran difference is \tSectDiffJF\,$^\circ$C and \sSectDiffJF\ RMS, and is
largest in the upper few hundred metres at high latitudes, where the two runs diverge
fastest through the mixed layer and the ice edge; the deep ocean contributes almost
nothing.

\subsection{Multidecadal drift}

\begin{figure*}[t]
\includegraphics[width=\textwidth]{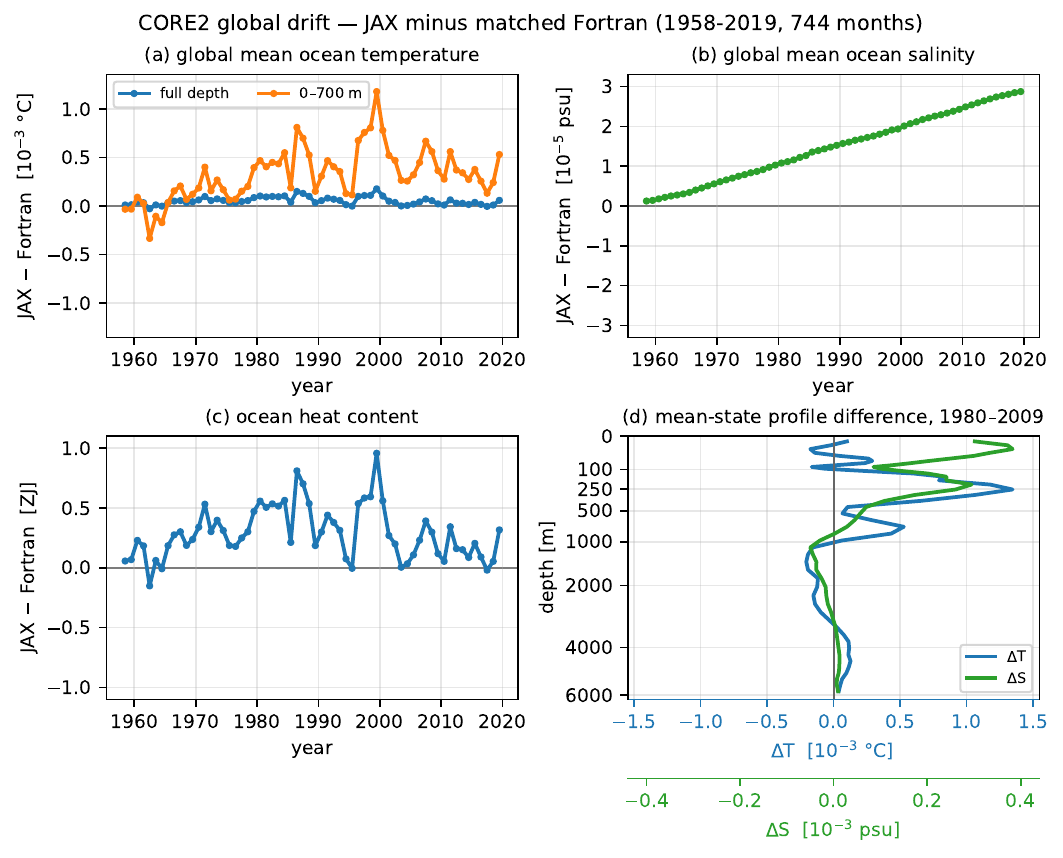}
\caption{Global drift over the CORE2 hindcast (\hindcastYears), shown throughout as
JAX \emph{minus} the matched Fortran run. (a)~volume-mean temperature, full depth and
0--700\,m; (b)~volume-mean salinity; (c)~ocean heat content; all three as annual means.
(d)~the JAX$-$Fortran temperature and salinity profiles averaged over \meanWindow, on a
sqrt-warped depth axis, each with its own $x$-axis. Panel~(c) is
panel~(a)'s full-depth curve rescaled, since the heat-content diagnostic uses a constant
$\rho_0 c_p$ and the fixed mesh volume.}
\label{fig:drift}
\end{figure*}

Figure~\ref{fig:drift} shows the difference between the two models' global integrals
over the hindcast. Both models cool from the PHC initial state and settle without
runaway drift, the volume-mean temperature falling from \tbarStart\,$^\circ$C to
\tbarEndPair, a change of \tbarOwnDrift\,$^\circ$C over six decades. The two codes
differ by \tbarDiff\,$^\circ$C, and that difference fluctuates about zero rather than
growing. The 0--700\,m mean diverges faster than the deep ocean and carries the larger
residual, but stays within \tbarSevenDiffMax\,$^\circ$C of the Fortran run even in its
most extreme single month. Ocean heat content stays within \ohcDiffMax\,ZJ
throughout and ends \ohcDiffEnd\,ZJ apart.

Volume-mean salinity changes by \sbarDriftJax\ on the practical salinity scale over
\hindcastYears, against \sbarDriftFor\ for Fortran. Salinity is the one integral whose
JAX$-$Fortran residual has a consistent sign, growing smoothly to \sbarDiffEnd\ by the
end of the hindcast rather than fluctuating about zero as the temperature difference
does. The
vertical structure of the difference, averaged over \meanWindow\ in panel~(d), reaches
\tProfDiffMax\,$^\circ$C at \tProfDiffDepth\,m and \sProfDiffMax\ at
\sProfDiffDepth\,m and decays towards the abyss, with the same upper-ocean confinement
as the sections of Fig.~\ref{fig:sections}.

A comparison of this length is what makes such residuals visible at all: an earlier
version of the port leaked freshwater through the global freshwater-flux normalization
of the $z^\star$ coordinate, a defect that showed up only as a slow salinity drift over
decades and that a shorter run would have missed (Sect.~\ref{sec:verification}).

\subsection{Sea ice}
\label{sec:seaice}

\begin{figure*}[t]
\includegraphics[width=\textwidth]{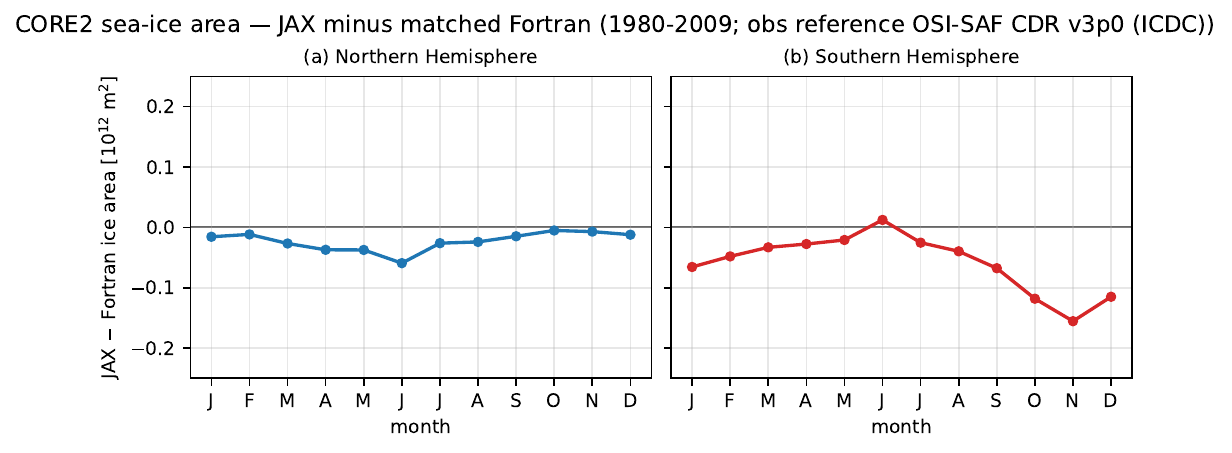}
\caption{CORE2 sea ice (\meanWindow): the seasonal cycle of the ice-area difference,
JAX \emph{minus} the matched Fortran run, for the Northern (a) and Southern (b)
Hemisphere.}
\label{fig:seaice}
\end{figure*}

Figure~\ref{fig:seaice} shows the seasonal cycle of the JAX$-$Fortran ice-area
difference. The largest monthly difference is \nhIceDiffMax\ $\times10^{12}$\,m$^2$ in
the Arctic (\nhIceDiffMonth) and \shIceDiffMax\ in the Antarctic (\shIceDiffMonth), or
\nhIceDiffPct\,\% and \shIceDiffPct\,\% of the ice area in those months; the larger
Antarctic value falls near the summer minimum, when the ice area itself is small. At the
four seasonal extremes the two models differ by at most \iceExtremeGap\
$\times10^{12}$\,m$^2$, and they share the same errors against OSI-SAF
\citep{osisaf2022cdr}: both carry too much Arctic ice through the cycle, with a seasonal
RMS difference of \nhRMSD\ $\times10^{12}$\,m$^2$, and both reproduce the
Southern-Hemisphere cycle to \shRMSD\ $\times10^{12}$\,m$^2$ with a low summer minimum.
The agreement holds at the minima as well as the maxima.

\subsection{Towards high resolution}

\begin{figure*}[t]
\includegraphics[width=12cm]{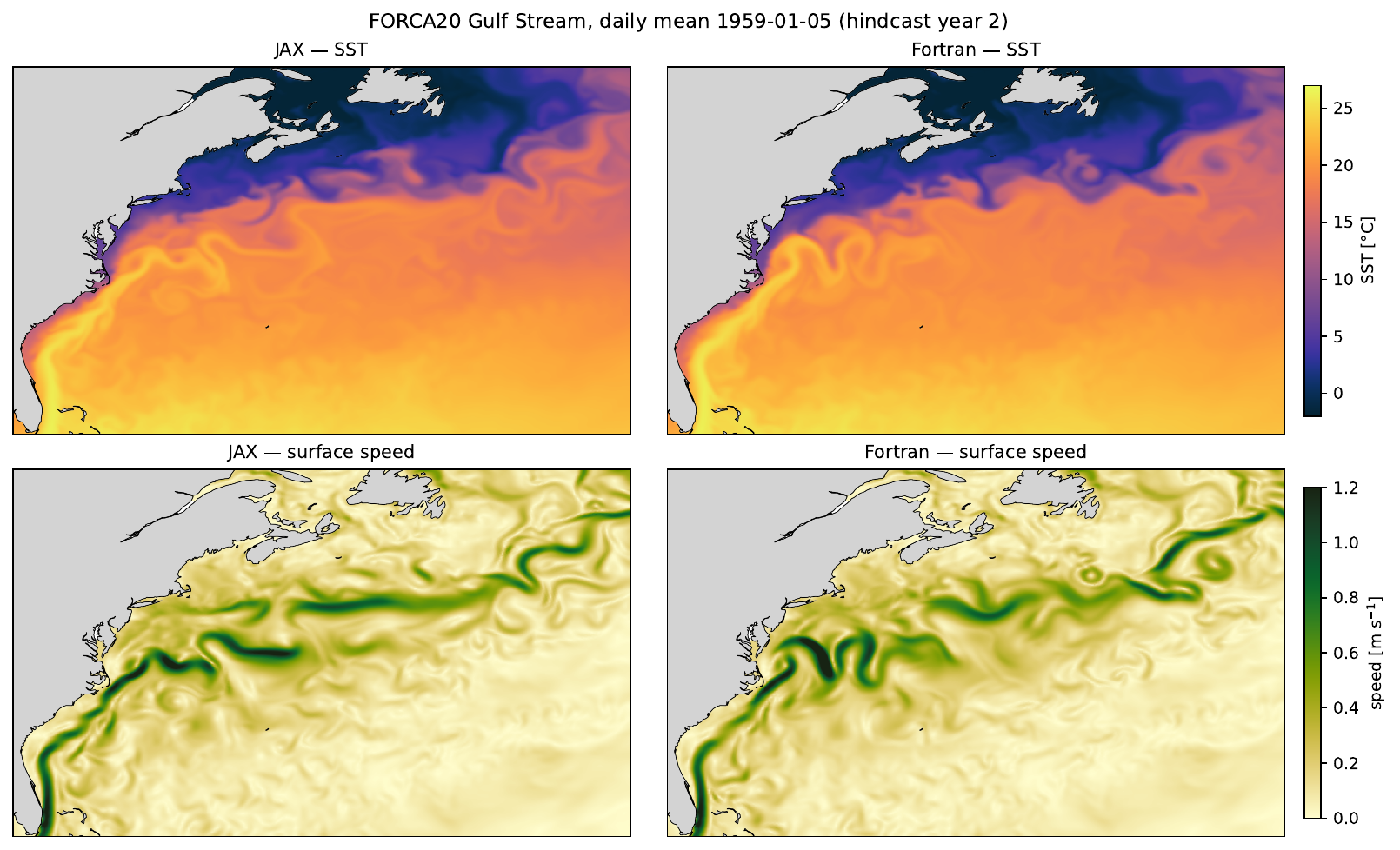}
\caption{Daily-mean SST (top) and surface speed (bottom) in the Gulf Stream region on
5~January of hindcast year~2 on the FORCA20 mesh ($2.13\times10^{6}$ surface vertices,
$\sim$14\,km local resolution): JAX (left) versus the matched Fortran run (right). The
Fortran fields come from a one-month re-integration of the reference run's January
1959, restarted from its end-of-1958 state with daily output enabled (the reference
run stores yearly means). The two models decorrelate during the preceding year of
eddying integration, so eddy positions differ by construction.}
\label{fig:gulfstream}
\end{figure*}

The test above is at coarse resolution, but the unstructured mesh is what makes FESOM2
attractive at high resolution, and the JAX port runs the same physics on eddy-permitting
meshes without modification. Figure~\ref{fig:gulfstream} compares the Gulf Stream region
in the second year of a hindcast on FORCA20 (Table~\ref{tab:meshes}). Both
implementations produce an eddying ocean of the same character, with a Gulf Stream
separating at Cape Hatteras, meanders and rings along the front, and comparable jet
sharpness and surface speed. A quantitative eddying
evaluation, including spectra, eddy statistics and boundary-current transports, is left
to future work.

\section{Performance and scalability}
\label{sec:performance}

FESOM2-JAX runs the whole mesh hierarchy of Table~\ref{tab:meshes} at useful speed, on
device counts from one GPU to 256, and from the same source on two different
accelerators. What limits it is the shape of the scaling curve rather than the absolute
throughput: the model is communication-bound, and it becomes so at partition sizes at
which the Fortran original is still scaling. We quantify both parts of that statement
below. Figure~\ref{fig:scaling} shows strong scaling and throughput across the
hierarchy, Fig.~\ref{fig:efficiency} reduces those curves to the single quantity that
governs them, and Table~\ref{tab:perf} gives representative production numbers.

\subsection{What is measured, and how}
\label{sec:perf-protocol}
All numbers are for the complete ocean--sea-ice model in each mesh's \emph{production}
configuration, with the same physics as the hindcasts of Sect.~\ref{sec:fidelity}:
$z^\star$ vertical coordinate, cvmix-TKE mixing, mEVP sea ice with its 120-iteration
subcycle, and, on CORE2, the only mesh coarse enough to need it, GM/Redi. The model is
driven by real JRA55-do forcing from the PHC initial state, and throughput is always
reported at the production timestep of Table~\ref{tab:meshes}. Measurements were made on
two GPU systems: the DKRZ Levante GPU partition, with four NVIDIA A100-80\,GB per node,
and the JUPITER booster at the J\"ulich Supercomputing Centre, with four NVIDIA GH200
Grace--Hopper superchips per node, 96\,GB of HBM3 each, and NDR200 InfiniBand. Unless a
number is labelled GH200, it was measured on A100. Each point runs on its fastest halo
transport (Table~\ref{tab:halo}). Timing follows a compile-once protocol: the step chain
is compiled, invoked once to warm it, and the same executable is then invoked a second
time and timed over 150 steps with device synchronization, twice per point, so that XLA
compilation never contaminates the per-step numbers. Compiling the full step chain
takes roughly half a minute on CORE2 and DARS and about two minutes for NG5 at 32
GPUs; the executable is cached on disk and reused across the chunks of a long run, so
it is paid once per configuration rather than once per job. In every run the temperature and velocity fields remain free of NaNs over the
timed window, so no timing comes from a diverged integration.

The protocol has two departures. DARS is \emph{benchmarked} at $\Delta t =
120$\,s rather than at the production 240\,s, because a cold-started benchmark window is
not stable at the production step on every partition of this mesh; the per-step cost is
timestep-independent, so the measured step time is unaffected. The DARS curve also starts at
16 GPUs: at 8 GPUs the production-physics step exceeds a practical XLA compilation
budget, an upstream compiler pathology at very large per-device partitions that is under
investigation. The same configuration compiles in about a minute from 16 GPUs upward.

\subsection{Absolute throughput}
The model is fast in absolute terms, and the whole hierarchy is affordable. A single
A100 runs the complete CORE2 configuration at \coreStepOneGPU\,s per step
(\coreSYPDOneGPU\ simulated years per wall-clock day, SYPD) within 11.2\,GiB of device
memory, so that configuration fits on one accelerator; a single node of four brings it
to \coreStep\,s (\coreSYPD\ SYPD) at \coreParEffFour\,\% parallel efficiency, which
makes the 62-year hindcast of Sect.~\ref{sec:fidelity} a modest computation, and a node
of four GH200s runs it at \ghStepCoreFour\,s (\ghSypdCoreFour\ SYPD). At the other end
of the hierarchy, NG5 sustains \ngSYPD\ SYPD on 64 A100s and \ghSypdNgSixtyFour\ SYPD on
64 GH200s. Table~\ref{tab:perf} gives the same numbers for every mesh.

\begin{figure*}[t]
\includegraphics[width=\textwidth]{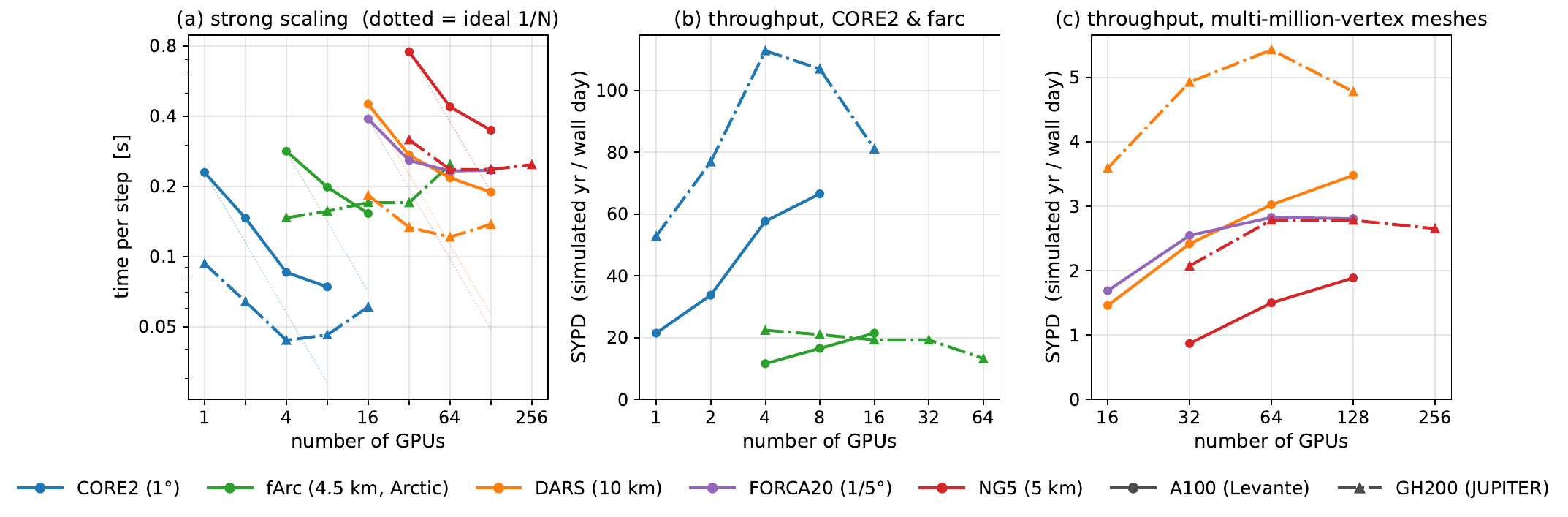}
\caption{Strong scaling of FESOM2-JAX for the full ocean$+$sea-ice model, using the same
source code on two GPU systems: A100 (DKRZ Levante; solid, circles) and GH200 (JUPITER
booster; dash-dotted, triangles). (a) Time per step versus the number of GPUs, per mesh;
dotted lines are ideal $1/N$ scaling anchored at each mesh's first A100 point. Every
point uses its fastest halo transport (Table~\ref{tab:halo}): on A100 the padded or
ragged all-to-all at 32 GPUs and below and the coloured \texttt{ppermute} at 64 and
above, on GH200 the coloured transport at every multi-node point. (b, c) Throughput in
simulated years per wall-clock day at each mesh's stable production timestep, split by
mesh size so that the small meshes' throughput (b) does not compress the
multi-million-vertex meshes' (c). The small CORE2 mesh saturates beyond one node on
either machine while the larger meshes strong-scale sub-linearly. GH200 is roughly twice
as fast per GPU wherever the per-device partition is large, and each mesh's knee
correspondingly moves to about half the device count. FORCA20 was measured on A100
only.}
\label{fig:scaling}
\end{figure*}

\begin{table}[t]
\caption{FESOM2-JAX at the production timestep on both GPU systems, run with the same source, JAX version, physics and protocol on each. FORCA20 was measured on A100 only. \emph{Speed-up} is the A100 per-step time divided by the GH200 one at the same GPU count. Peak device memory is a property of the partition rather than of the accelerator (NG5 at 32 GPUs measures 48.5\,GiB on GH200 against the 48.2\,GiB tabulated here).}
\label{tab:perf}
\begin{tabular}{lrrrrrrrrr}
\tophline
 &  & \multicolumn{4}{c}{A100 (Levante)} & \multicolumn{4}{c}{GH200 (JUPITER)} \\
\cline{3-6}\cline{7-10}
Mesh (\#GPU) & vertices & s/step & SYPD & GPU-h/yr & GiB/GPU & s/step & SYPD & GPU-h/yr & speed-up \\
\middlehline
CORE2 (4) & 127\,k & 0.085 & 57.7 & 2 & 3.1 & 0.044 & 112.8 & 1 & 1.96 \\
CORE2 (8) & 127\,k & 0.074 & 66.5 & 3 & 1.7 & 0.046 & 106.9 & 2 & 1.61 \\
DARS (16) & 3.16M & 0.450 & 1.5 & 263 & 28.1 & 0.183 & 3.6 & 107 & 2.46 \\
FORCA20 (16) & 2.13M & 0.389 & 1.7 & 227 & 21.6 & -- & -- & -- & -- \\
NG5 (32) & 7.40M & 0.755 & 0.9 & 883 & 48.2 & 0.317 & 2.1 & 370 & 2.38 \\
NG5 (64) & 7.40M & 0.438 & 1.5 & 1024 & 34.2 & 0.236 & 2.8 & 551 & 1.86 \\
\bottomhline
\end{tabular}
\belowtable{}
\end{table}

\subsection{One quantity governs the scaling}
\label{sec:perf-collapse}
We measure the work a device performs in \emph{vertex-levels}: one layer of the water
column beneath one surface vertex, so a mesh of $N$ surface vertices and $L$ layers has
$N \times L$ of them, from \meshCoreLanes\ million on CORE2 to \meshNgLanes\ million on
NG5 (Table~\ref{tab:meshes}). The model holds its state in dense arrays of that shape
and advances every entry each timestep, including the masked ones below the sea floor,
so vertex-levels per second is a rate of work comparable across meshes of different size
and depth.

\begin{figure*}[t]
\includegraphics[width=\textwidth]{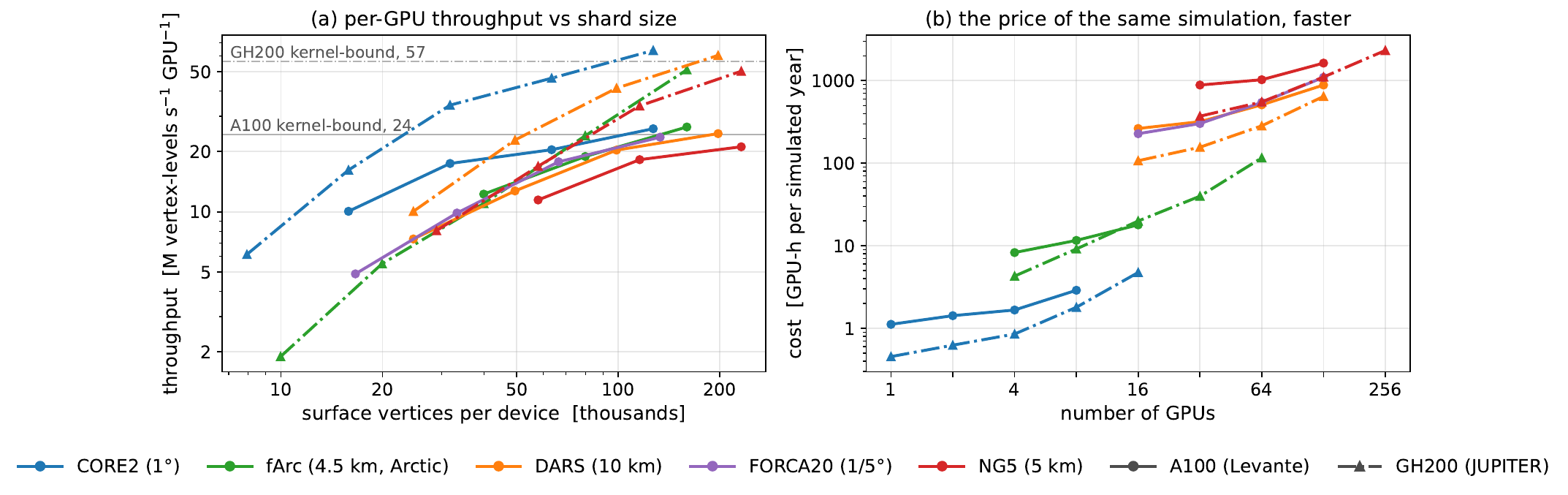}
\caption{The same measurements as Fig.~\ref{fig:scaling}, reduced to what governs them.
(a) The rate at which one GPU advances vertex-levels, against the number of the mesh's
surface vertices it owns, taken as the mesh's vertex count divided by the device count
(the FESOM2 partitions used here are balanced to within 1.2\,\% at every point plotted);
grey guides mark the kernel-bound plateau. CORE2 lies above the common curve at small
shards because it is the only mesh carrying GM/Redi, so that the same halo is overlapped
with more arithmetic. (b) Accelerator hours per simulated year against device count.}
\label{fig:efficiency}
\end{figure*}

Counted this way, the rate one GPU sustains depends on how many surface vertices it owns
and on little else (Fig.~\ref{fig:efficiency}a). Above \thrPlateauVerts\ surface vertices
per device the rate is \thrPlateau\ million vertex-levels per second on an A100 and
\thrPlateauGH\ million on a GH200, and the five meshes agree to
$\pm\thrPlateauSpread$\,\% (\thrPlateauLo--\thrPlateauHi\ million) although they differ
by a factor of \meshSizeSpan\ in size, run from a $1^\circ$-equivalent configuration to
$\sim$5\,km, and carry between 47 and 69 layers. Below that threshold the rate falls away
steadily: half as many vertices gives \thrHalfKeep\,\% of the plateau, and fewer than
30\,000 vertices only \thrSmallShard\,\%. The halo is a rim around a two-dimensional
subdomain, so the communication-to-work ratio grows as the inverse square root of the
vertices per device and is independent of the layer count, which is why meshes of 47 and
69 layers fall on one curve.

A mesh therefore strong-scales while its shards are large and saturates once they are
not, so the device count at which it stops improving grows with its size, and a faster
accelerator raises the throughput at a given shard size without changing the shard size
at which scaling ends.

\subsection{Strong scaling}
\label{sec:perf-strong}
CORE2 exhausts its parallelism within two nodes: the second node still reduces the step
time, but at 58\,\% doubling efficiency, and at 8 GPUs each device holds about
16\,000 vertices, the small-shard end of Fig.~\ref{fig:efficiency}a.
The eddy-permitting meshes strong-scale sub-linearly and keep improving out to the
largest A100 partition we tested: the first doubling of each retains
\forcaEffThirtyTwo--\ngEffSixtyFour\,\% parallel efficiency, and DARS still returns
\darsEffSixtyFour\,\% on its second (32$\to$64 GPUs). At 128 GPUs a DARS step is
\darsStepOneTwentyEight\,s and an NG5 step \ngStepOneTwentyEight\,s, both still falling,
while FORCA20, the smallest of the three, has flattened at
\forcaStepOneTwentyEight\,s. fArc, between these two regimes, gains a factor 1.85 from
4 to 16 GPUs.

Whether these meshes keep scaling at high device counts also depends on the halo
transport. With a single $P$-way all-to-all, DARS and FORCA20 \emph{turn over} past 64
GPUs, because that transport's per-device message count grows with the device count
$P$. The coloured \texttt{ppermute}, whose cost is set by the mesh's colouring rather
than by $P$, removes the turnover. Section~\ref{sec:perf-halo} shows how their ranking
changes with scale.

\subsection{The same code on a second machine}
\label{sec:perf-portability}
Because the model is ordinary JAX code, running it on a different accelerator required
no change to the source: the JUPITER installation is a virtual environment built from
the same \texttt{jax[cuda12]} wheels, which are available for the machine's aarch64
hosts, with mesh exports regenerated on site and the same YAML run configurations, and
needed no container image or vendor-specific build. We repeated the campaign there from
a single GPU to 256 (dash-dotted curves in Fig.~\ref{fig:scaling}; the GH200 columns of
Table~\ref{tab:perf}). The gain is confined to throughput per GPU, where it is a factor
of \thrPlateauRatio\ at the kernel-bound end of Fig.~\ref{fig:efficiency}a; it decays as
the partitions shrink and communication becomes the limiting cost, and at the smallest
partitions it inverts, fArc on 16 GPUs being \emph{slower} on GH200 than on A100
(\ghRatioFarcSixteen$\times$). These kernels are gather- and bandwidth-bound rather than
FLOP-bound, so the ratio follows the memory bandwidth of the two devices, a factor of
about two, far more closely than their arithmetic peak, a factor of about three and a
half. The range over which the model scales is unchanged: every mesh turns over at
roughly half the device count it does on A100, NG5 for instance saturating at 64 GPUs
(\ghStepNgSixtyFour\,s), unchanged at 128 (\ghStepNgOneTwentyEight\,s) and slightly
slower at 256 (\ghStepNgTwoFiftySix\,s). This follows from the per-GPU gain: a device
twice as fast completes a shard's work in half the time while the communication per step
is unchanged, so the compute--communication crossover arrives at half the shard count,
and a faster machine is best used with fewer, larger shards. The multi-node points were
measured on a busy shared system, where repeats within one allocation agree to
0.05\,\% but the spread between allocations is $\sim 5$\,\%, so the turnover locations
carry that uncertainty.

\paragraph{Two lessons about portability.} A source that transfers unchanged does not
carry its tuning with it, in two respects. The first is that the ranking of the halo
transports depends on the machine as well as on the mesh, and the Levante ranking does
not transfer. On GH200, the padded all-to-all is slower than the coloured
\texttt{ppermute} at every multi-node point we measured, by factors between 1.1 and 5.5,
and its timings do not reproduce between allocations; the coloured transport reproduces
and is fastest at every point beyond one node, so every multi-node GH200 point uses it.
Contention from our own concurrent jobs, a degraded fabric (a direct bandwidth probe
through the same communication library found it healthy), and partition imbalance were
each tested and rejected as explanations, and the cause remains open. Because the four
transports are interchangeable behind a run-time flag, adapting to the new machine
amounted to selecting a different one.

The second is that a code whose inner loops are written by a compiler also inherits the
compiler's judgement about how to write them, including its mistakes. At the NG5
partition into 128 devices, and at no other partition, XLA merged the global
surface-flux balance sum into the sea-ice thermodynamics, which it compiles as a single
block of 1425 operations, and switched that block to a slower code path, producing one
383\,ms kernel and a 2.7$\times$ increase in step time. The effect was bracketed by
unaffected points at 64 and 256 GPUs, reproduced on separate allocations, and present
under all three transports. Forcing the compiler to keep the sum separate, which leaves
the outputs bit-identical, removes it and reduces the step time from 0.637 to
\ghStepNgOneTwentyEight\,s; that barrier is now the default and the point plotted in
Fig.~\ref{fig:scaling} is the corrected one. This kind of failure belongs to
compiler-generated code, and is part of the price of the portability demonstrated above.

\subsection{Where the time goes}
Switching components off decomposes the CORE2 4-GPU production step. The bare ocean
core, comprising dynamics, the free surface, tracers, and forcing,
accounts for $\sim 58$\,ms per step. The mEVP sea ice adds $\sim 14$\,ms, almost
entirely from the 120 subcycled rheology iterations, each of which performs its own halo
exchange. The GM/Redi eddy parameterization adds $\sim 9$\,ms, the $z^\star$ moving
vertical coordinate $\sim 5$\,ms, and the TKE mixing closure only $\sim 3$\,ms.
The single-GPU baseline prices the communication directly: relative to ideal scaling
from \coreStepOneGPU\,s, communication and duplicated halo work account for about
\commShareFour\,\% of the measured 4-GPU step. Both point at the halo exchange, which
lies on the critical path of every stencil and every sea-ice subcycle.

\subsection{Which transport is fastest}
\label{sec:perf-halo}

The ranking of the four transports of Sect.~\ref{sec:halo} changes with scale
(Table~\ref{tab:halo}). On the small CORE2 mesh, the minimal-volume
ragged and padded exchanges are fastest and the coloured transport is slowest: the SSH
conjugate-gradient solve fires a few hundred small two-dimensional exchanges per step,
these are latency-bound, and $K$ sequential \texttt{ppermute} rounds pay $K$ launch
latencies where a single all-to-all pays one. On the largest configuration, NG5 at 64
GPUs, the order reverses. The three-dimensional tracer exchanges dominate, these are
bandwidth-bound, and the coloured transport's flat-in-$P$ volume beats both the ragged
all-to-all (by 8\,\%) and the padded one (by 37\,\%). DARS at 32 GPUs sits just before
the crossover: ragged and padded tie at 0.35\,s and the coloured transport remains
slightly slower at 0.39\,s. An exact adjoint therefore costs nothing: at both ends of
the mesh range the fastest transport with an exact adjoint is faster than the
forward-only ragged all-to-all (Table~\ref{tab:halo}), and the large end is where
gradients are scientifically useful.

Choosing the transport per point recovers part of the communication cost but not all of
it. The unstructured partition makes the exchanges irregular, since each rank
communicates with a data-dependent set of neighbours, and their relative cost grows as
the subdomains shrink, which is the effect Fig.~\ref{fig:efficiency}a measures. Two
improvements are already reflected in Fig.~\ref{fig:scaling}: the padded and coloured
transports move each stencil's halo in one statically shaped collective, and fusing the
sea-ice subcycle's velocity pair and the solver's paired reductions removed a further
$\sim 9$\,\% of the CORE2 8-GPU step.

\begin{table}[t]
\caption{The four halo-exchange transports, and measured per-step time (full model,
uniform benchmark configuration of the transport study, so that the ranking rather than
the absolute times is the object of comparison) at the two ends of the mesh range.
``Adjoint'' is whether reverse-mode is exact; $P$ is the device count. The all-gather is
impractical at NG5 (it gathers the full $\sim$4\,GB field onto every device). The ragged
all-to-all is forward-only because its transpose is defective in JAX 10.1.}
\label{tab:halo}
\begin{tabular}{lllrr}
\tophline
Transport & Adjoint & Per-device volume & CORE2/4 & NG5/64 \\
\middlehline
all-gather broadcast    & exact        & whole field            & 0.090 & --- \\
ragged all-to-all       & forward-only & halo only, $P$-way     & 0.086 & 0.589 \\
padded all-to-all       & exact        & halo $\times\,P$ slots & 0.080 & 0.742 \\
coloured \texttt{ppermute} & exact     & halo $\times\,K$ rounds & 0.093 & 0.542 \\
\bottomhline
\end{tabular}
\belowtable{}
\end{table}

\subsection{Optional accelerations}
Two further optimizations are implemented as opt-in switches and are OFF in every number
reported here, because each changes results within solver or floating-point tolerance
rather than bit for bit. The first computes the per-step atmospheric forcing on the
device, removing host-side interpolation and transfer. The second replaces the
preconditioner of the sea-surface-height solve with a degree-3 Chebyshev polynomial,
which reaches the same residual in roughly one third of the iterations (127 to 42 on the
CORE2 operator); because the adjoint of the solve is obtained by implicit
differentiation, gradients are unchanged to $10^{-13}$. With both on, annual means of a
one-year integration deviate from the production configuration by no more than repeated
baseline runs deviate from one another, while the end-to-end production loop accelerates
by 21--53\,\%, the gain growing with mesh size because the host-forcing and solver shares
both grow. On NG5 the on-device forcing path currently costs more device time than it
saves on the host, a compiler-optimization pathology under investigation, so only the
Chebyshev preconditioner is worth enabling there ($-4$ to $-5$\,\%).

\subsection{The cost of a gradient}
Reverse-mode differentiation has a measurable cost: modest in time and larger in memory.
On one A100, with per-step checkpointing, a reverse-mode step of the complete model,
defined here as the \texttt{value\_and\_grad} of a scalar diagnostic over an $N$-step
window including the checkpointed recomputation of the forward pass, costs 0.92\,s,
compared with 0.196\,s for the forward step. This is a factor of 4.7 and is stable
between $N=12$ and $N=24$ windows. Compiling the reverse program is a one-time cost of
about six minutes. Memory, rather than time, is the binding constraint on the window:
the backward pass stores one model state per step, measured at 0.83\,GiB per step on
CORE2 on top of a $\sim$19\,GiB working set (29\,GiB at $N=12$, 39\,GiB at $N=24$).
Day-scale adjoint windows therefore fit on one 80-GB device, while longer windows use
the two-level checkpointing of Sect.~\ref{sec:jax-implementation}, trading recomputation
for storage.

\subsection{CPU execution and its scope}
The same source also runs on CPUs. This is a convenience rather than a performance
claim: it supports development, continuous integration, teaching, and laptop-scale
experimentation. On one
128-core node, the full CORE2 model executes at \coreStepCPU\,s per step (\coreSYPDCPU\
SYPD) as a single process parallelized by XLA's thread pool, with no partition files.
Launched as sixteen processes of eight cores each under the \texttt{shard\_map} layer,
with the coloured halo transport, it runs at 1.10\,s per step (4.5 SYPD), which also
exercises the sharded code path, including gradients, without a GPU. It scales across
nodes: DARS takes 16.4\,s per step on eight nodes and 8.6\,s on sixteen, a speed-up of
$1.91\times$. NG5 does not run on CPU at all, because every process builds the global
model state before taking its share and so needs close to the whole-model footprint
whatever the process count. The gap to the Fortran original has two parts. First, the
individual kernels run about $2.7\times$ slower, the price an array framework pays for
generating code from whole-array operations rather than from hand-written loops. Veros
reports $1.0$--$1.4\times$ against its own Fortran ancestor \citep{hafner2021veros}.
Second, parallel
efficiency is about $2.5\times$ worse, and the cause is not the halo exchange but the
global sums: the CPU all-reduce grows linearly with process count where MPI's grows
logarithmically, and the surface-elevation solve needs two per iteration, so adding
processes stops reducing the time per step beyond about 32. The Fortran original runs 512 ranks and reaches 87.8 SYPD
on four CPU nodes, faster than any of our GPU configurations. Many-CPU production
therefore belongs to the Fortran original, while the JAX shadow targets GPU execution,
gradient-based applications, and the rapid implementation of new model capabilities.

\section{Outlook: differentiability}
\label{sec:outlook}

The capability that most clearly distinguishes the shadow from the Fortran model it
shadows is differentiability. Because the entire time loop, including ocean dynamics,
tracer transport, vertical mixing, and sea-ice rheology, is written as pure JAX
functions, the gradient of any scalar diagnostic with respect to any input, boundary
condition, or model parameter is available by reverse-mode automatic differentiation
through the full integration, at a cost independent of the number of parameters and
measured in Sect.~\ref{sec:performance}.

\begin{figure}[t]
\includegraphics[width=8.3cm]{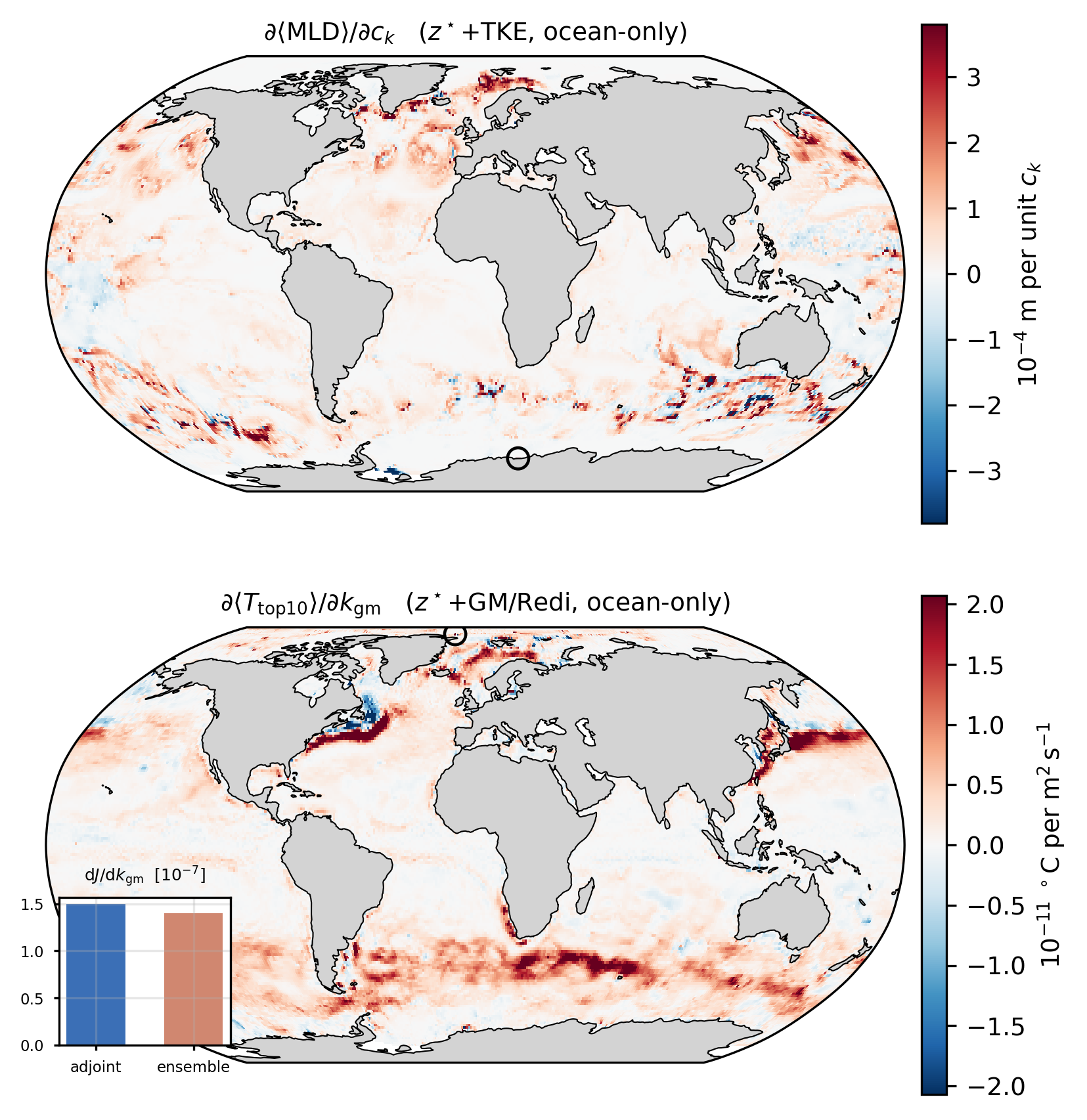}
\caption{Adjoint sensitivity maps computed by reverse-mode automatic differentiation
through FESOM2-JAX on the CORE2 mesh. Top: gradient of the global-mean mixed-layer
depth with respect to a vertex-wise field of the TKE mixing parameter $c_k$; bottom:
gradient of upper-ocean (top ten layers) mean temperature with respect to the
Gent--McWilliams/Redi coefficient $k_{\mathrm{gm}}$. Each map is one backward pass over
a six-hour window from the PHC initial state under JRA55-do forcing. Circles mark the
vertices at which the map value is verified against a pointwise finite difference; the
inset compares the scalar $k_{\mathrm{gm}}$ gradient, the sum of the map, with the
forward-ensemble estimate.}
\label{fig:sensitivity}
\end{figure}

Figure~\ref{fig:sensitivity} illustrates this with two parameter-sensitivity maps on the
CORE2 mesh. In each, a scalar parameter is promoted to a field with one value per
surface vertex, so that a single reverse-mode pass returns the complete map
$\partial J/\partial\theta(x)$; the same map by finite differences would need one
forward integration per vertex, here $1.3\times 10^5$ runs. The top panel differentiates
the global-mean mixed-layer depth with respect to the TKE mixing parameter $c_k$, the
bottom the mean temperature of the upper ten layers with respect to the
Gent--McWilliams/Redi coefficient $k_{\mathrm{gm}}$. Both are ocean-only, with sea ice
off and only the differentiated parameterization active. Differentiation through the sea
ice, including the 120-iteration mEVP subcycle, is available, but the iterated explicit
rheology amplifies reverse-mode signals over the window, so our gradient applications to
date hold the ice fixed; sea ice has required comparable care in the adjoint of the MIT
general circulation model \citep{heimbach2010seaice}.

The sum of each map matches a central finite
difference of the forward model to relative $6\times10^{-7}$, a pointwise finite
difference reproduces the local value at the most sensitive vertex, and the
$k_{\mathrm{gm}}$ gradient agrees to within 7\,\% with an independent twelve-member
forward-ensemble estimate. Component-by-component gradient checks run in the
continuous-integration suite (Sect.~\ref{sec:verification}).

Forward-mode differentiation of the same source yields the tangent-linear model,
which needs no stored trajectory: its memory does not grow with the window length.
The two modes are complementary: the adjoint returns the sensitivity of one scalar
diagnostic to a field of parameters at a cost independent of the field's size, while
the tangent-linear model returns the response of every model field, at every point,
to one scalar parameter, at a cost independent of the number of outputs. Because the
two are transposes of the same linearization, the same scalar derivative can be
computed both ways. In half-day windows on the CORE2 mesh, with the full physics
active and the ice held fixed in the derivative, the two routes agree to better than
1\,\%.

Two caveats bound the capability. A gradient through chaotic dynamics is informative
only over finite horizons, beyond which sensitivities grow exponentially
\citep{lea2000sensitivity,kohl2002adjoint}, so the six-hour maps of
Fig.~\ref{fig:sensitivity} are instantaneous sensitivities rather than an equilibrium
response. And the model contains switches: sea-ice concentration thresholds, the
triggers in the mixing schemes, and the flux limiter in tracer advection. The gradient
is defined everywhere except at those switching points, which is why the
finite-difference checks of Sect.~\ref{sec:verification} are taken at states away from
them. What the capability enables, gradient-based calibration against
observations and hybrid models whose trainable components are trained through the
dynamical core, is ongoing work and lies beyond the scope of this paper.

\conclusions
\label{sec:conclusions}

We have presented FESOM2-JAX, a re-implementation of the unstructured-mesh
ocean--sea-ice model FESOM2 in Python and JAX. The Fortran original is a production
code: it is the ocean component of the AWI climate model for CMIP7 and runs in
kilometre-scale initiatives including the Destination Earth Climate Digital Twin.
FESOM2-JAX is its code shadow, a projection of the same model onto Python and its
machine-learning ecosystem, kept demonstrably faithful to the original that casts it,
and we have evaluated it against the three requirements such a shadow must meet.

It is faithful. In a \hindcastYears\ hindcast at $1^\circ$-equivalent resolution, with
matched physics and forcing, the mean states of the two codes differ from each
other by two orders of magnitude less than either differs from observations, and the
runs agree for six decades in temperature, heat content and the sea-ice seasonal
cycle.

It is usable and it is fast. The model is a NumPy-style codebase that a student can
read and modify; a run is specified by a single YAML file; restarts and output do not
depend on how many devices produced them; and a regression test fails on any change
that alters the shipped configuration's results. The same source runs, unchanged,
from a laptop CPU to 256 GPUs. The complete model at $1^\circ$ fits on a single
GPU, and a node of four modern GPUs (GH200 superchips) integrates \ghSypdCoreFour\
simulated years per wall-clock day, so the hindcast is about half a day of computation; meshes of up to 7.4 million surface
vertices, about 5\,km, scale to 128 GPUs. What limits the model is communication rather
than arithmetic. The small meshes stop gaining beyond one node; the large ones scale
only because the halo exchange is implemented so that its cost per device does not
grow with the number of devices.

What the shadow adds to the original is the gradient, a capability that has been hard
for ocean models to acquire and maintain, and one the Fortran original never had. Because the whole time loop is differentiable,
a single backward pass returns the sensitivity of a model diagnostic to a parameter at
every one of the mesh's $1.3\times10^{5}$ surface vertices at once, where the same map
by finite differences would take one integration per vertex. Differentiated forward,
the same loop yields the tangent-linear model: the response of every field to one
parameter, with no stored trajectory. Both derivatives come from the source that
defines the forward model and remain consistent with it as the code evolves.

Among models written natively in differentiable frameworks, FESOM2-JAX is to our
knowledge the first global ocean--sea-ice model of CMIP-class complexity, and
the first on an unstructured mesh. What this paper demonstrates is the capability
itself, on short-window sensitivities verified against finite differences.
The applications it points to are potential rather than demonstrated. The same
machinery underpins ocean state estimation and variational data assimilation. Model
parameters could be calibrated against observations by
gradient descent, through the tangent-linear model when the tuned coefficients are few
and through the adjoint when they are a field. Sensitivity maps computed with respect
to the ocean state rather than to a parameter show where an observation would most
constrain a chosen diagnostic, the starting point of observing-system design. Because
the derivative extends to
whatever is added to the code, neural-network components could be trained through the
ocean physics rather than beside it. Realizing them is ongoing work.

A production model ordinarily faces a choice between its validated code and the
platforms where new methods are developed. Code shadows
avoid that choice: the Fortran model remains the single source of truth; its shadows
(a C++/Kokkos translation for performance portability, \citealp{koldunov2026llmport},
and FESOM2-JAX for experimentation and gradients) project it onto new technology stacks
and are verified by running alongside it; and what a shadow produces (tuned parameter
sets, trained machine-learning components, validated numerical variants) returns to the
production model. Through its shadows, a Fortran model meets modern hardware, machine
learning, and a new generation of developers without a full rewrite.

\codeavailability{FESOM2-JAX is developed openly at
\url{https://github.com/koldunovn/fesom_jax}; the exact version used here will be
tagged and archived at Zenodo on acceptance (DOI:~\texttt{TODO}).
The reduction and plotting scripts that produce every figure in this paper from the
derived data will be archived together with the code on acceptance; they depend only
on standard scientific-Python libraries.}

\dataavailability{The model is forced by JRA55-do \citep{tsujino2018jra55do}. The
observational references are the PHC3.0 hydrography \citep{steele2001phc}, which is also
the initial condition, and the OSI-SAF sea-ice concentration climate-data record. A
runnable data package for the CORE2 configuration (the mesh, the derived PHC initial
state, the domain decompositions, and one year of JRA55-do forcing) is archived at
Zenodo (\url{https://doi.org/10.5281/zenodo.21324319}) and retrieved by a fetch script
shipped with the code. The
small derived arrays that produce every figure will be shared on acceptance, archived
together with the scripts; the raw model output (the CORE2 and FORCA20 hindcasts and
the matched Fortran runs) is several terabytes and is available from the authors on
request.}

\authorcontribution{TODO: author contributions to be completed.}

\competinginterests{The authors declare that they have no conflict of interest.}

\begin{acknowledgements}
We thank the FESOM2 development team. Computations were performed on the DKRZ Levante
supercomputer and, for the GH200 scaling campaign of Sect.~\ref{sec:performance}, on the
JUPITER supercomputer at the J\"ulich Supercomputing Centre.
\end{acknowledgements}

\financialsupport{This work was primarily supported through the core funding of the
Alfred Wegener Institute, Helmholtz Centre for Polar and Marine Research (AWI), within
the Helmholtz Association. Additional support for individual authors was provided
through the following projects and funding sources. The contribution by NK, IK, DP and
SL was supported by the TerraDT (Digital Twin of Earth system for Cryosphere, Land
surface and related interactions) project, which has received funding from the
European Union's Horizon Europe research and innovation programme under Grant
Agreement no.~101187992. SD, PS and NK were supported by projects M5, S2 and S1 of the
Collaborative Research Centre TRR181 ``Energy Transfer in Atmosphere and Ocean''
funded by the Deutsche Forschungsgemeinschaft (DFG, German Research Foundation) --
Projektnummer 274762653. SC is funded by the WarmWorld Better project of the German
Federal Ministry of Research, Technology and Space under the funding code 01LK2202A;
the responsibility for the content of this publication lies with the authors. The work
of SB has been supported by the European Union's Destination Earth Initiative and
relates to tasks entrusted by the European Union to the European Centre for
Medium-Range Weather Forecasts implementing part of this Initiative with funding by
the European Union. Views and opinions expressed are those of the authors only and do
not necessarily reflect those of the European Union or the European Commission.
Neither the European Union nor the European Commission can be held responsible for
them. AK was supported by the HClimRep project, funded by the Helmholtz Foundation
Model Initiative (HFMI). TJ was supported by the EERIE project (Grant Agreement
No~101081383) funded by the European Union. Views and opinions expressed are however
those of the author(s) only and do not necessarily reflect those of the European Union
or the European Climate Infrastructure and Environment Executive Agency (CINEA).
Neither the European Union nor the granting authority can be held responsible for
them.}

\appendix
\section{Component-by-component description of the ported model}
\label{app:components}

This appendix collects the description of the ported components, so that the model can
be read without consulting \citet{danilov2017fesom2} and
\citet{scholz2019fesom2,scholz2022fesom2}. Numerical settings quoted here are those
used in the present study and are summarized in Table~\ref{tab:components}.

\subsection{Continuous equations and horizontal discretization}

The ocean component solves the hydrostatic, Boussinesq primitive equations.

The horizontal mesh is an unstructured triangulation, and the discretization is cell-vertex finite-volume. Scalar fields, including temperature, salinity, pressure, and sea-surface height, are located at mesh vertices and advanced on the control volumes, while horizontal velocity is located at triangle centroids \citep{danilov2017fesom2}. Horizontal operators are therefore indirect: divergences, gradients, and fluxes are assembled by gathering values along precomputed vertex, edge, and element neighbor lists and scattering edge-wise antisymmetric contributions back. The same gather/scatter stencil structure, rather than structured-grid array shifts, gives the model its smoothly variable resolution and also accounts for much of its computational cost (Sect.~\ref{sec:performance}). Lateral boundaries are no-slip.

\subsection{Vertical coordinate}

The vertical coordinate is the arbitrary Lagrangian--Eulerian (ALE) coordinate of FESOM2, used throughout this study in its $z^\star$ form \citep{adcroft2004rescaled}. In this formulation, the free-surface displacement is distributed uniformly over the stretchable part of the water column, so layer thicknesses vary with the change in sea-surface height while layer interfaces retain their nominal stacking. Two consequences are important for what follows. First, surface freshwater exchange is represented as a real volume flux, not as a virtual salt flux, so the global freshwater budget must close exactly. The salinity-affecting bug in the python JAX port, discussed in Sect.~\ref{sec:fidelity}, violated precisely this closure. Second, the pressure-gradient force must be evaluated on the time-varying geometry. FESOM2-JAX therefore uses the density-Jacobian formulation of \citet{shchepetkin2003pgf}, as the original does for $z^\star$. The linear free surface option is also available.

\subsection{Equation of state}

In-situ
density and hydrostatic pressure are computed from the polynomial equation of state of
\citet{jackett1995eos}; the thermal-expansion and haline-contraction coefficients that
feed the mixing and eddy parameterizations follow \citet{mcdougall1987neutral}, and
the Brunt--V\"ais\"al\"a frequency is smoothed once over the vertex neighbourhood, as in
the original.

\subsection{Time stepping and the free surface}

One model timestep advances momentum, the free surface, and the tracers in sequence,
with the sea-ice component executed first (Sect.~\ref{sec:seaice-model}). Momentum
uses a second-order Adams--Bashforth estimate, with FESOM2's stabilizing offset
$\epsilon=0.1$, for the Coriolis and advection terms; momentum advection is evaluated
in flux form on the scalar control volumes. Sea-surface height is advanced by treating
the external mode semi-implicitly, with two implicitness weights in the notation of
\citet{danilov2017fesom2}: $\alpha$ on the transport divergence in the elevation
equation, and $\theta$ on the surface elevation in the momentum equation. Setting both
to $1/2$ recovers the Crank--Nicolson scheme; we use FESOM2's default
$\alpha=\theta=1$, which is fully implicit. The resulting elliptic system is
re-assembled on each time step under the full free surface and is static under the
linear free surface. It is solved at every step by a preconditioned
conjugate-gradient (CG) iteration, with the symmetric preconditioner of
\citet{marshall1997mitgcm} that FESOM2 uses, to a relative residual of $10^{-5}$. The
solve is warm-started from the previous step solution and typically converges in a few
iterations.

Layer thicknesses are then updated from the new surface height ($z^\star$), and the
quasi-vertical diasurface velocity is diagnosed by integrating the horizontal transport divergence
upward from the no-flux bottom. Vertical viscosity and diffusion are treated
implicitly with tridiagonal solves in each column, removing the vertical CFL
restriction from mixing. FESOM2's \texttt{w\textunderscore split}, which moves the part
of the vertical advective velocity that exceeds the explicit CFL limit into the implicit
solve, is also ported. It is normally enabled on high-resolution meshes to improve
stability, but none of the runs reported here use it. The CORE2 configuration of
Sect.~\ref{sec:fidelity} runs at $\Delta t = 1800,\mathrm{s}$.

\subsection{Interior physics and surface fluxes}

Tracers are advected with the third-order upwind MUSCL-type scheme \citep{danilov2017fesom2} in the horizontal direction and the fourth-order centered scheme in the vertical direction, limited by the flux-corrected-transport (FCT) scheme of \citet{zalesak1979fct}: a first-order upwind solution is corrected by
limited antidiffusive fluxes from the high-order estimate. The advected field enters the fluxes through the same Adams--Bashforth interpolation as the Coriolis and momentum advection terms.

Vertical mixing uses the prognostic
turbulent-kinetic-energy closure of \citet{gaspar1990tke} in the CVMix formulation
adopted by FESOM2 \citep[cvmix-TKE;][]{griffies2015cvmix,vanroekel2018kpp,scholz2022fesom2}: one TKE
equation per column, integrated implicitly, with the standard constants
($c_k=0.1$, $c_\epsilon=0.7$) and background TKE value and viscosity of
$10^{-5}$ and $10^{-4}$\,m$^2$\,s$^{-1}$ respectively. The K-profile parameterization
\citep{large1994kpp} and the Richardson-number scheme of \citet{pacanowski1981pp} are
also ported and configuration-selectable; all runs in this paper use cvmix-TKE.

Where the mesh does not resolve eddies, eddy effects
are parameterized by Gent--McWilliams bolus advection \citep{gent1990isopycnal} and
Redi isoneutral diffusion \citep{redi1982isopycnal}. As in FESOM2, the bolus transport
is obtained by solving a vertical boundary-value problem for the eddy streamfunction
in each column \citep{ferrari2010bvp} rather than from local slopes alone; the
transfer coefficient (here capped at 1000\,m$^2$\,s$^{-1}$) is scaled with the local
horizontal resolution and attenuated in the vertical, and neutral slopes are tapered
where they steepen. The parameterization is a pure diagnostic of the current density
field, recomputed every step. It is active in the CORE2 hindcast and switched off in
the configurations used for the performance measurements of
Sect.~\ref{sec:performance}. Long integrations on fArc, which is coarse outside its
refined region, also use it.

Momentum dissipation uses FESOM2's biharmonic filter, assembled edge-wise in two passes of an approximate Laplacian with a coefficient set by
the local mesh scale and the velocity difference across the edge, so that dissipation concentrates where the flow is rough and relaxes to a small background value where it is
smooth. The operator is that of \citet{danilov2017fesom2}, iterated to biharmonic form; the flow-aware coefficient is the one introduced by \citet{juricke2020kinematic}, who
note that, being based on velocity differences, it is reminiscent of the Smagorinsky viscosity \citep{smagorinsky1963}.

The model is driven by the JRA55-do reanalysis dataset \citep{tsujino2018jra55do}: of three-hourly fields (10-m winds, air temperature, specific humidity, downward short- and longwave radiation, rain, and snow). They are interpolated bilinearly to the mesh and linearly in time, and turbulent fluxes are computed interactively with the NCAR bulk formulae \citep{large2009bulk}, whose Monin--Obukhov stability iteration runs a fixed number of passes (a choice made
for differentiability, Sect.~\ref{sec:jax-implementation}). Shortwave radiation penetrates the water column with a chlorophyll-dependent attenuation profile \citep{sweeney2005swpene}. River
runoff is a climatological field, and sea-surface salinity is restored weakly to a
monthly climatology with a piston velocity of 10\,m per 60 days.

\subsection{Sea ice}
\label{sec:seaice-model}

Sea ice is FESOM2's single-class dynamic--thermodynamic model, based on the
Finite-Element Sea Ice Model \citep[FESIM;][]{danilov2015fesim} and discretized at the
same mesh vertices as the ocean scalars. Its prognostic variables are ice concentration, ice
and snow volume per unit area, and ice velocity at the vertices.

The momentum balance uses the viscous--plastic rheology with an
elliptical yield curve (aspect ratio 2) and the \citet{hibler1979seaice} strength
$P = P^\ast h\, e^{-C(1-A)}$ (here $P^\ast = 3\times10^4$\,N\,m$^{-2}$, $C=20$),
solved with the modified elastic--viscous--plastic (mEVP) iteration \citep{bouillon2013mevp,kimmritz2015mevp}, the scheme FESOM2 uses in production for its favourable convergence and cost \citep{koldunov2019fastevp}. Each ocean timestep runs a fixed 120 pseudotime iterations with stabilization parameters $\alpha=\beta=250$; each iteration evaluates element strain rates and stresses,
scatters the stress divergence to the vertices, and solves the pointwise implicit momentum
update including Coriolis and an ice--ocean drag (coefficient $5.5\times10^{-3}$). Standard EVP is retained as a configuration option.

The three ice tracers are advected with a Taylor--Galerkin finite-element FCT scheme of \citet{lohner1987fct}, the sea-ice
finite-element counterpart of the ocean FCT. Thermodynamics is the zero-layer scheme of \citet{semtner1976thermo}: a fixed-iteration Newton solve for the ice skin
temperature, growth rates averaged over seven ice-thickness classes, lead closing following \citet{hibler1979seaice}, snow-to-ice conversion by flooding, and a salinity-dependent freezing point.

The ice component runs first within each timestep: it receives
the ocean surface state (temperature, salinity, surface currents, and the sea-surface
tilt), executes dynamics, advection, and thermodynamics, and returns to the ocean the
ice-modulated surface stress, the net heat flux, the shortwave flux gated by ice
cover, and the surface salinity forcing. Under $z^\star$ the ice--ocean freshwater
exchange is a real volume flux with an explicit salt flux for the brine content, and
the global-mean freshwater flux is removed each step so that the free surface
conserves volume --- the budget closure whose violation produced the salinity drift
dissected in Sect.~\ref{sec:fidelity}.

\subsection{Nominal and area-weighted mesh resolution}
\label{app:meshes}

The nominal resolution of Table~\ref{tab:meshes} is a design figure, not a uniform one:
a mesh built to concentrate resolution is coarse over most of its area. On fArc, which
is CORE2 outside the Arctic, four-fifths of the vertices lie in the 4.5\,km refined
region, yet the area-weighted mean resolution over the global ocean is
\meshFarcResArea\,km. \citet{koldunov2026lec} compare the latitude-based and
eddy-refined mesh families directly and find that eddy-refined meshes reproduce global
energetics comparable to latitude-based ones at lower cost.

\noappendix

\bibliographystyle{copernicus}
\bibliography{references}

\end{document}